\documentclass[twocolumn,notitlepage,prb,showpacs,preprintnumbers,superscriptaddress,amsmath,amssymb,citeautoscript]{revtex4-2}
\pdfoutput=1
\usepackage{graphicx}
\usepackage[caption=false]{subfig}
\usepackage{epstopdf}
\usepackage{dcolumn}
\usepackage{color}
\usepackage{amsmath,amssymb}
\usepackage[english]{babel}
\usepackage{bm}

\usepackage[hyperindex]{hyperref}
\hypersetup{colorlinks,citecolor=blue, filecolor=blue,linkcolor=blue,urlcolor=blue}
\usepackage[section]{placeins}

\usepackage{multirow}
\usepackage{array}

\newcommand{\bpm}{\begin{pmatrix}}
\newcommand{\epm}{\end{pmatrix}}
\newcommand{\bs}{\boldsymbol}
\newcommand{\be}{\begin{equation}}
\newcommand{\ee}{\end{equation}}
\newcommand{\beq}{\begin{eqnarray}}
\newcommand{\eeq}{\end{eqnarray}}
\newcommand{\angstrom}{\mbox{\normalfont\AA}}

\DeclareMathOperator{\im}{Im}

\DeclareMathOperator{\tr}{tr}

\begin{document}

\title{Surface states and quasiparticle interference in Bernal and rhombohedral graphite with and without trigonal warping}

\author{Vardan Kaladzhyan}
\email{vardan.kaladzhyan@phystech.edu}
\affiliation{Department of Physics, University of Basel, Klingelbergstrasse 82, CH-4056 Basel, Switzerland}
\author{Sarah Pinon}
\affiliation{Institut de Physique Th\'eorique, Universit\'e Paris Saclay, CEA
CNRS, Orme des Merisiers, 91190 Gif-sur-Yvette Cedex, France}
\author{Fr\'ed\'eric Joucken}
\affiliation{Department of Physics, University of California, Santa Cruz, California 95064, USA}
\author{Zhehao Ge}
\affiliation{Department of Physics, University of California, Santa Cruz, California 95064, USA}
\author{Eberth A. Quezada-Lopez}
\affiliation{Department of Physics, University of California, Santa Cruz, California 95064, USA}
\author{T. Taniguchi}
\affiliation{International Center for Materials Nanoarchitectonics, National Institute for Materials Science, 1-1 Namiki, Tsukuba 305-0044, Japan}
\author{K. Watanabe}
\affiliation{Research Center for Functional Materials, National Institute for Materials Science, 1-1 Namiki, Tsukuba 305-0044, Japan}
\author{Jairo Velasco Jr}
\affiliation{Department of Physics, University of California, Santa Cruz, California 95064, USA}
\author{Cristina Bena}
\affiliation{Institut de Physique Th\'eorique, Universit\'e Paris Saclay, CEA
CNRS, Orme des Merisiers, 91190 Gif-sur-Yvette Cedex, France}

\date{\today}

\begin{abstract}
We use an exact analytical technique [Phys. Rev. B \textbf{101}, 115405 (2020), Phys. Rev. B \textbf{102}, 165117 (2020)] to recover the surface Green's functions for Bernal (ABA) and rhombohedral (ABC) graphite. For rhombohedral graphite we recover the predicted surface flat bands. For Bernal graphite we find that the surface state spectral function is similar to the bilayer one, but the trigonal warping effects are enhanced, and the surface quasiparticles have a much shorter lifetime. We subsequently use the T-matrix formalism to study the quasiparticle interference patterns generated on the surface of semi-infinite ABA and ABC graphite in the presence of impurity scattering.  We compare our predictions to experimental STM data of impurity-localized states on the surface of Bernal graphite which appear to be in a good agreement with our calculations. 
\end{abstract}

\maketitle

\section{Introduction}

\begin{figure}[t]
	\centering
	\includegraphics[width=1.1\linewidth]{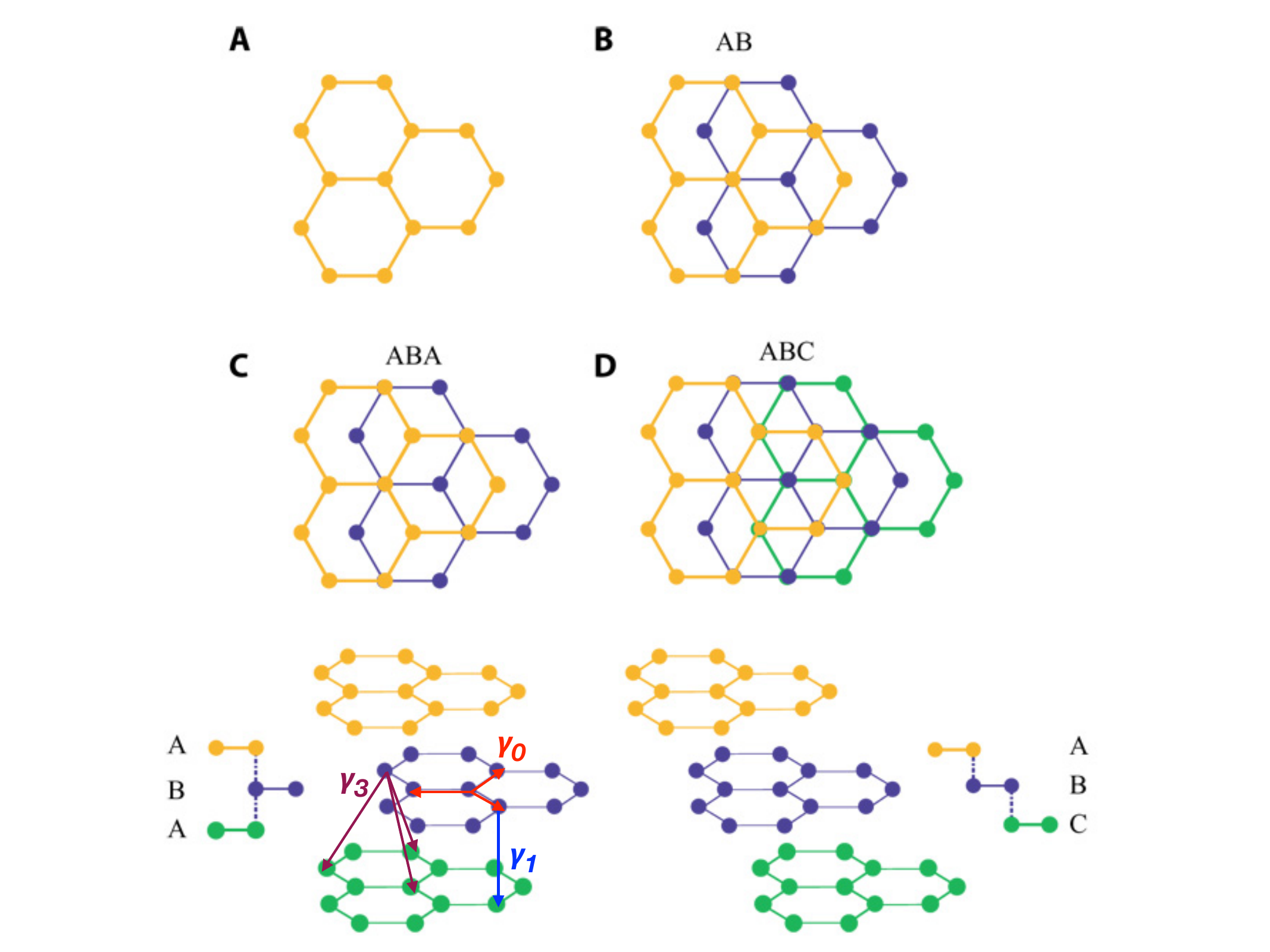}
	\caption{Lattice structure for  A: single-layer graphene, B: bilayer graphene with AB stacking, C: Bernal-stacked graphite, D: rhombohedral-stacked graphite. Modified from Ref.~[\onlinecite{Shan2018}].}
	\label{fig:grapheneStacking}
\end{figure}

Despite graphite having been known to theorists for decades \cite{Wallace1947}, it was not until after the experimental discovery of graphene in 2004 \cite{Novoselov2004,Novoselov2005} that graphene-based systems started to attract a lot of attention, both in experimental and theoretical groups in condensed matter physics. While a lot of interest is devoted to few-layer systems since these are cleaner and possibly more interesting for nanoelectronics applications, here we focus rather on 3D graphite, in particular on the relationship between the electronic properties of its surface states, accessible via ARPES or STM, and the properties of few-layer systems. We consider two types of stacking: the Bernal stacking (also known as ABA) and the rhombohedral stacking (also known as ABC). Both are shown in Fig.~\ref{fig:grapheneStacking}.  To describe the surface states we use the technique presented in Refs.~[\onlinecite{Pinon2020a}] and [\onlinecite{Pinon2020b}], in which the surface Green's functions are obtained by introducing an infinite-strength plane-like impurity, which effectively cuts the system in two, and by solving exactly the problem using the T-matrix formalism. 

Thus, for semi-infinite ABC graphite we recover the existence of the surface state flat bands previously predicted to appear in multi-layered ABC graphene \cite{Yelgel2016,Zhang2010,Kopnin2011} and observed in Ref.~[\onlinecite{Henni2016}], and we determine their extent in a fully semi-infinite system.  For the ABA graphite the surface states are similar to those of bilayer graphene in that they have a parabolic dispersion close to the Dirac point. However, the effects of the trigonal warping are more pronounced for graphite surface states than for a bilayer system. Moreover, we find that the quasiparticle lifetimes and coherence lengths are greatly reduced for the surface states, making these states less sharp in momentum space.

We subsequently consider the quasiparticle interference (QPI) patterns arising on the surface of graphite from impurity-scattering processes. These features have been studied before in connection to various aspects of graphene physics \cite{Cheianov2006,Mariani2007,Kivelson2003,Bena2005,Bena2008,Bena2009a,Bena2016,Wehling2007,Peres2006,Peres2007,Vozmediano2005,Ando2006, Pogorelov2006,Skrypnyk2006,Skrypnyk2007,Katsnelson2007,Dutreix2013,OjedaCollado2015,Dutreix2016,Dutreix2019}.  We demonstrate that the oscillations in the local density of states associated with the impurity scattering decay much faster on the surface of graphite, be it ABA or ABC, than in bilayer graphene(BLG) and trilayer graphene (TLG). Furthermore, we show that trigonal warping gives rise to a three-fold symmetry reflected both in the spectral functions and the quasiparticle interference patterns; this effect is enhanced on the surface of graphite compared to the BLG and gives rise to impurity states with a striking three-fold symmetry. Our theoretical results on graphite reproduce well experimental STM data obtained on graphite and thick graphene films.


The paper is organized as follows: In Sec.~\ref{sec:BernalImpurity} we present the tight-binding models used to describe graphene and graphite. In Sec.~\ref{sec:BandsfiniteGraphiteSlab} we compute the band structure for multilayer graphene systems using numerical tight-binding calculations; the latter can be used as a reference point for the analytical semi-infinite-system calculations in Sec.~\ref{sec:SpectralFunctionSemiInfGraphite}. Furthermore, in Sec.~\ref{sec:SpectralFunctionSemiInfGraphite} we calculate the surface spectral function for both ABA and ABC graphite. In Sec.~\ref{sec:QPIpatterns} we compute the QPI patterns (both in real space and in momentum space) for ABA and ABC graphite, BLG and TLG. In Sec.~\ref{sec:ExperimentalData}, we present experimental STM results obtained on freshly exfoliated graphite crystals, as well as on a multilayer graphene film ($\sim10$ layers thick) supported by hexagonal boron nitride (hBN), and we show that both the short localization length and the trigonal warping features agree well with our theoretical results. In Sec.~\ref{sec:EvolutionWithLayersNumber} we describe the evolution of the QPI main features with the number of layers by comparing the results for semi-infinite graphite with those for 2, 3, 4, and 8 layers.  We leave the conclusions to Sec.~\ref{sec:Conclusion}, while in the Appendix we present the models used in Sec.~\ref{sec:EvolutionWithLayersNumber} for multilayer systems.


\section{Tight-binding models for bilayer and trilayer graphene, as well as for ABA and ABC graphite}\label{sec:BernalImpurity}

To model BLG and TLG, as well as graphite, we consider three types of hopping: the intralayer nearest-neighbor hopping $\gamma_0$, the interlayer nearest-neighbor hopping $\gamma_1$, i.e., between two sites which have the same coordinates in the $(x,y)$ plane and belong to adjacent layers, and finally, the interlayer hopping denoted $\gamma_3$ responsible for trigonal warping        \cite{Charlier1991,McCann2013,Joucken2020a}. In what follows we will take $\gamma_0=3.3$eV, $\gamma_1=0.42$eV, and for $\gamma_3$  we will consider either $\gamma_3=0$ or $\gamma_3=-0.3$eV. If not mentioned all units will be considered to be given in eV.

For bilayer graphene the $\bs{k}$-space Hamiltonian written in the basis $\left\{ \psi_{\bs{k}}^{A1},\, \psi_{\bs{k}}^{B1},\, \psi_{\bs{k}}^{A2},\, \psi_{\bs{k}}^{B2} \right\}$, where $A, B$ and $1, 2$ refer to sublattices and layers, respectively, is given by:
\begin{equation}
 \mathcal{H}_{BLG}(\bs{k}) = 
\begin{pmatrix}
0 & h_0(\bs{k}) & 0 & h_3(\bs{k}) \\
h_0^*(\bs{k}) & 0 & \gamma_1 & 0 \\
0 & \gamma_1 & 0 & h_0(\bs{k}) \\
h_3^*(\bs{k}) & 0 & h_0^*(\bs{k}) & 0 
\end{pmatrix},
\label{hh1}
\end{equation}
where we defined 
\begin{align}
h_0(\bs{k}) &= - \gamma_0 \left[1 + 2 e^{- i \frac{3}{2} a_0 k_x} \cos \left(\frac{\sqrt{3}}{2} a_0 k_y \right) \right], \\
h_3(\bs{k}) &= - \gamma_3 \varepsilon,
\end{align}
and 
\begin{equation}
\varepsilon \equiv 2 e^{- i \frac{3}{2} a_0 k_x} \cos \left(\frac{\sqrt{3}}{2} a_0 k_y \right) + e^{-i3 a_0 k_x}.
\label{eq:epsilon}
\end{equation}


The trilayer graphene that we will consider in our calculations is the ABC (rhombohedral stacking), with the corresponding Hamiltonian  written in the basis $\left\{ \psi_{\bs{k}}^{A1},\, \psi_{\bs{k}}^{B1},\, \psi_{\bs{k}}^{A2},\, \psi_{\bs{k}}^{B2},\, \psi_{\bs{k}}^{A3},\, \psi_{\bs{k}}^{B3} \right\}$:
\begin{equation}
\nonumber \mathcal{H}_{TLG}(\bs{k}) = 
\begin{pmatrix}
0 & h_0(\bs{k}) & 0 & h_3(\bs{k}) & 0 & 0\\
h_0^*(\bs{k}) & 0 & \gamma_1 & 0 & 0 & 0 \\
0 & \gamma_1 & 0 & h_0(\bs{k}) & 0 & h_3(\bs{k}) \\
h_3^*(\bs{k}) & 0 & h_0^*(\bs{k}) & 0 & \gamma_1 & 0 \\
0 & 0 & 0 & \gamma_1 & 0 & h_0(\bs{k}) \\
0 & 0 & h_3^*(\bs{k}) & 0 & h_0^*(\bs{k}) & 0 
\end{pmatrix}.
\label{hh2}
\end{equation}

The Bernal-stacked (ABA) graphite (see Fig.~\ref{fig:grapheneStacking}) is the most common as well as the most stable form of graphite\cite{Nery2020}. 
The  Hamiltonian is given by:
\begin{eqnarray}
\label{eq:hamiltonianBernal}
&&\mathcal{H}_{ABA}(\bs{k}) = \sum_{l} h_0(\bs{k}) c_{\bs{k}, l, A}^\dagger c_{\bs{k}, l, B}  \times
 \\
&&\times \left[1 + 2 e^{- i \frac{3}{2} a_0 k_x} \cos \left(\frac{\sqrt{3}}{2} a_0 k_y \right) \right] \nonumber \\
&&+ \gamma_1 \sum_{l \text{ even}} c_{\bs{k}, l, B}^\dagger c_{\bs{k}, l+1, A} 
+ \gamma_1 \sum_{l \text{ odd}} c_{\bs{k}, l, A}^\dagger c_{\bs{k}, l+1, B}  \nonumber \\
&&+ \sum_{l \text{ even}} h_3(\bs{k}) c_{\bs{k}, l, A}^\dagger c_{\bs{k}, l+1, B} 
\nonumber \\
&&
+ \sum_{l \text{ odd}} h_3^*(\bs{k}) c_{\bs{k}, l, B}^\dagger c_{\bs{k}, l+1, A} + \text{H.c.}\nonumber,
\end{eqnarray}
where  $l$ labels the layer. We define a two-layer unit cell, so that the Hamiltonian in Eq.~(\ref{eq:hamiltonianBernal}) can be rewritten as:
\begin{eqnarray}
\mathcal{H}_{ABA}(\bs{k})& =&  \sum_{l = l_-, l_+} h_0(\bs{k}) c_{\bs{k}, l, A}^\dagger c_{\bs{k}, l, B}  \\
&&
\times \left[1 + 2 e^{- i \frac{3}{2} a_0 k_x} \cos \left(\frac{\sqrt{3}}{2} a_0 k_y \right) \right] \nonumber \\
&&+ \gamma_1 c_{\bs{k}, l_-, B}^\dagger c_{\bs{k}, l_+, A} \left(1 + e^{-2 i d_0 k_z} \right) \nonumber \\
&&+h_3(\bs{k}) c_{\bs{k}, l_-, A}^\dagger c_{\bs{k}, l_+, B}  \left(1 + e^{-2 i d_0 k_z} \right) + \text{H.c.} \nonumber, 
\label{eq:hamiltonianBernalBulk}
\end{eqnarray}
where $\bs{k} = (k_x, k_y, k_z)$, $l_-$ ($l_+$) corresponds to the lower (upper) layer of the unit cell, $d_0$ is the distance between two neighboring layers, and $\varepsilon$ is defined in Eq.~(\ref{eq:epsilon}). We take $a_0=1.42,\angstrom$, and $d_0 = 3.35\,\angstrom$. 

The tight-binding Hamiltonian for rhombohedral graphite is given by:
\begin{eqnarray}
&&\mathcal{H}_{ABC}(\bs{k}) =  \sum_{l} h_0(\bs{k}) c_{\bs{k}, l, A}^\dagger c_{\bs{k}, l, B} \times
 \\
&&\times \left[1 + 2 e^{- i \frac{3}{2} a_0 k_x} \cos \left(\frac{\sqrt{3}}{2} a_0 k_y \right) \right] \nonumber \\
&&+ \gamma_1 c_{\bs{k}, l, B}^\dagger c_{\bs{k}, l+1, A} +h_3(\bs{k}) c_{\bs{k}, l, A}^\dagger c_{\bs{k}, l+1, B} + \text{H.c.}\nonumber.
\label{eq:hamiltonianRhombo}
\end{eqnarray}
This can be rewritten as:
\begin{eqnarray}
&& \mathcal{H}_{ABC}(\bs{k}) =  \sum_{l = l_-, l_0, l_+} h_0(\bs{k}) c_{\bs{k}, l, A}^\dagger c_{\bs{k}, l, B} \times
\nonumber \\
&&\times \left[1 + 2 e^{- i \frac{3}{2} a_0 k_x} \cos \left(\frac{\sqrt{3}}{2} a_0 k_y \right) \right] \nonumber \\
&& + \gamma_1 c_{\bs{k}, l_-, B}^\dagger c_{\bs{k}, l_0, A} + \gamma_1 c_{\bs{k}, l_0, B}^\dagger c_{\bs{k}, l_+, A} \nonumber \\
&& + \gamma_1 c_{\bs{k}, l_+, B}^\dagger c_{\bs{k}, l_-, A} e^{3 i a_0 k_x} e^{3 i d_0 k_z} \nonumber \\
&& + h_3(\bs{k}) c_{\bs{k}, l_-, A}^\dagger c_{\bs{k}, l_0, B} 
+ h_3(\bs{k}) c_{\bs{k}, l_0, A}^\dagger c_{\bs{k}, l_+, B} \nonumber \\
&& + h_3(\bs{k}) c_{\bs{k}, l_+, A}^\dagger c_{\bs{k}, l_-, B} e^{3 i a_0 k_x} e^{3 i d_0 k_z} + \text{H.c.} \nonumber, 
\label{eq:hamiltonianRhomboBulk}
\end{eqnarray}
where $l_-, l_0, l_+$ correspond to the lower, middle and upper layers of the unit cell respectively. The unit cell is thus composed of six sites: there is an $A$-sublattice and a $B$-sublattice site for each of the three layers.

Using these Hamiltonians, we can define the unperturbed Matsubara Green's function:
\begin{equation}
G_0(\bs{k}, i \omega_n) = \left[ i \omega_n \mathbb{I} - \mathcal{H}(\bs{k}) \right]^{-1}.
\label{eq:MatsubaraGreen}
\end{equation}

To simplify notations hereinafter we will assume that $a_0$ is set to unity, and hence $d_0=2.36$ and all momenta are given in units of $1/a_0$.

\section{Band structure for a finite graphite slab}\label{sec:BandsfiniteGraphiteSlab}
\begin{figure}[h]
	\centering
	\hspace*{-1.3cm}
	\includegraphics[width=0.6\linewidth]{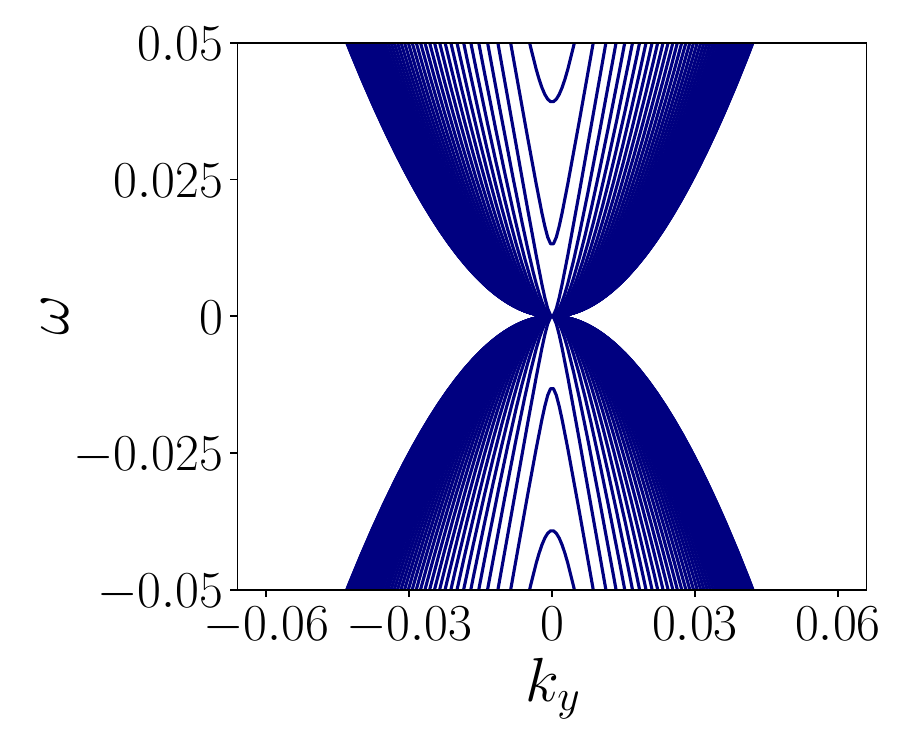}
	\caption{Spectrum for Bernal-stacked graphite around the $K$ point for 100 layers, with $\gamma_3=0$.}	
	\label{fig:graphiteBernalKpoint}
\end{figure}

We now use numerical tight-binding calculations in order to obtain the energy spectrum for a finite slab of graphite. We first plot in Fig.~\ref{fig:graphiteBernalKpoint} the spectrum around the $K$ point for a 100-layer ABA graphite slab with no trigonal warping ($\gamma_3 = 0$). As expected, in full accordance with Ref.~[\onlinecite{Wallace1947}], the spectrum shows multiple bands with parabolic dispersion.

\begin{figure}[h]
	\centering
	\hspace*{-0.4cm}
		\includegraphics[width=0.32\columnwidth]{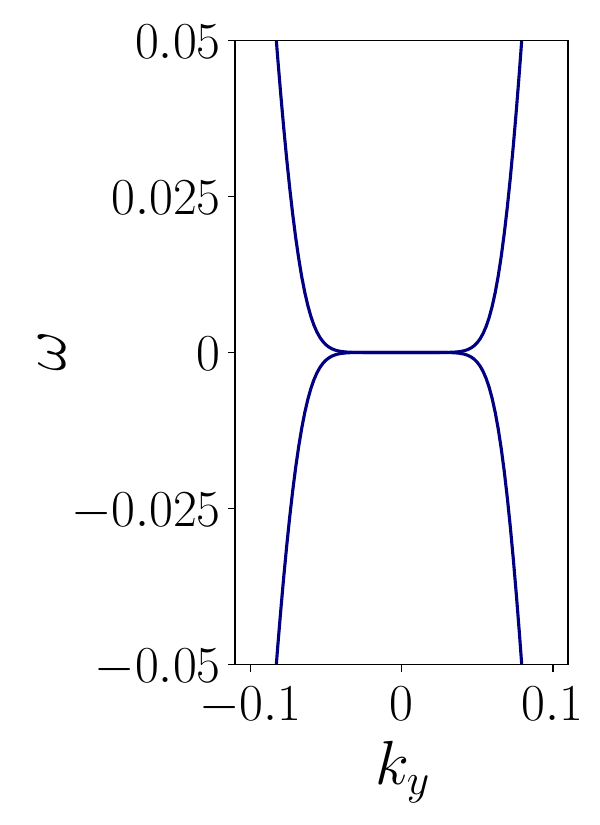}
      	\includegraphics[width=0.32\columnwidth]{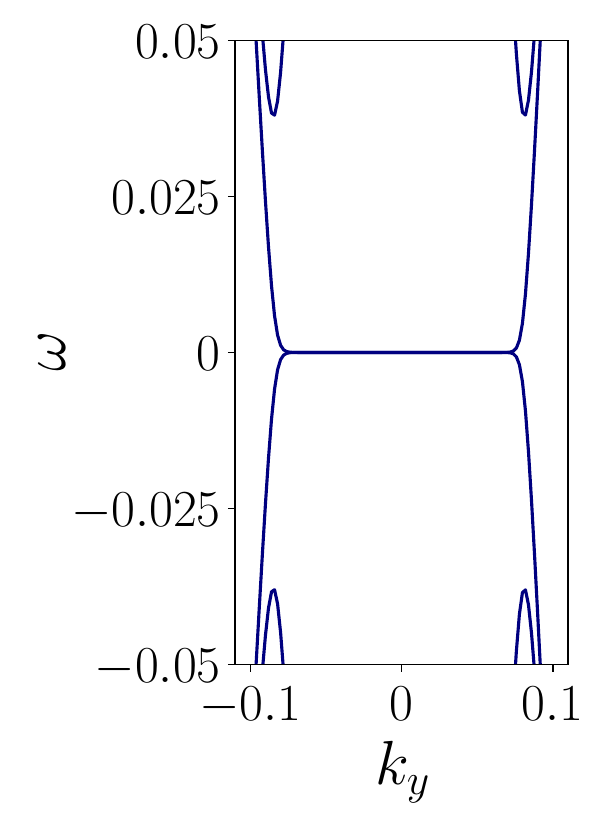}
      	\includegraphics[width=0.32\columnwidth]{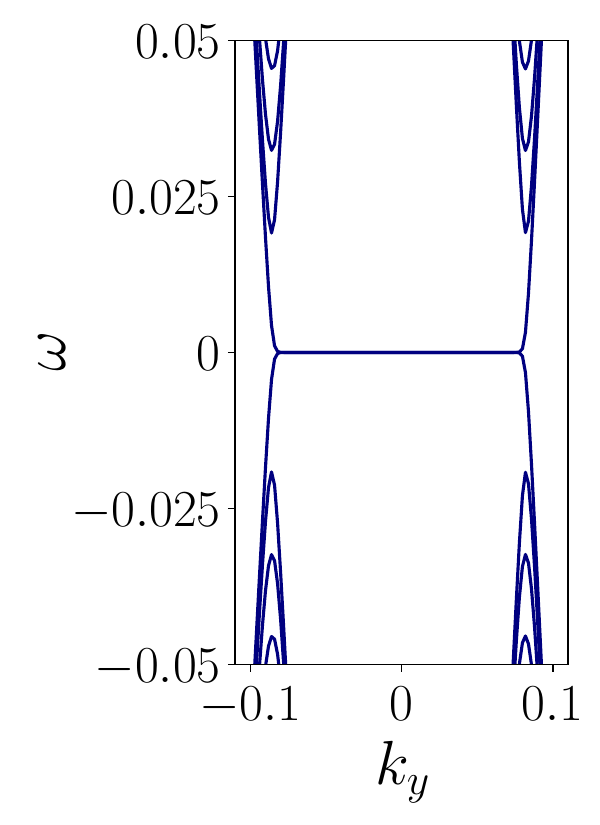}
	\caption{Spectrum for ABC graphite around the $K$ point for 10, 50 and 100 layers (from left to right), and  $\gamma_3=0$.}
	\label{fig:graphiteRhomboKpoint}
\end{figure}

We then calculate the spectrum of ABC graphite in the absence of trigonal warping. This is plotted in Fig.~\ref{fig:graphiteRhomboKpoint} for different numbers of layers. As we can see, the surface states in this case form a flat band at zero energy, which extends further away from the $K$ point when increasing the number of layers. It seems that when the number of layers tends to infinity, the extension of the flat band reaches a limit value in $\bs{k}$ space.

These tight-binding calculations will serve as a reference for comparison with the analytical calculations for semi-infinite graphite in the next section. Direct and indirect experimental evidences of the existence of the above mentioned surface states can be found in Ref.~[\onlinecite{Yin2019}] and [\onlinecite{Shi2020}]. 

\begin{figure}[h]
	\centering
	\includegraphics[width=0.7\linewidth]{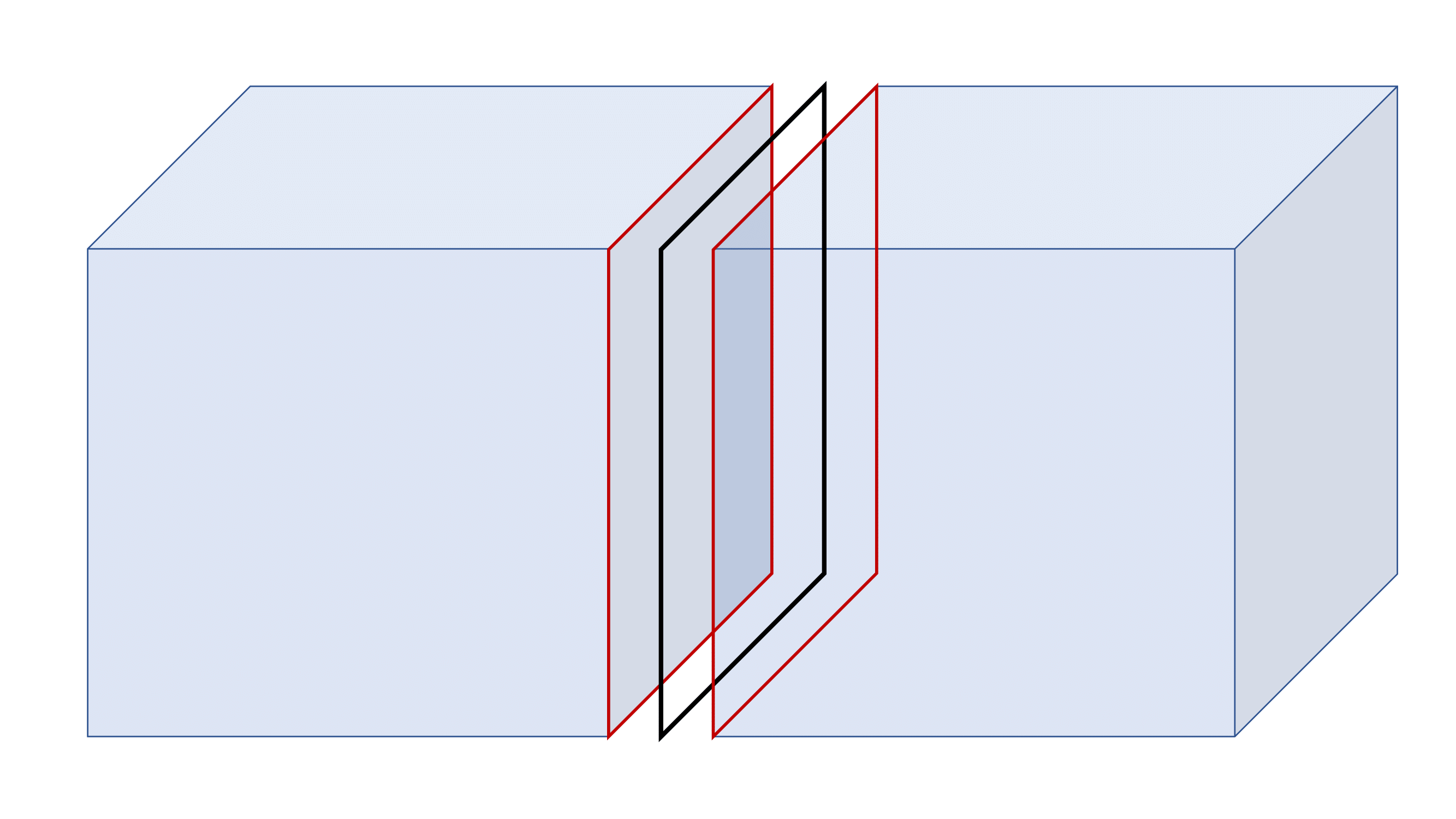}
	\caption{A plane-like impurity (in black) separating a 3D infinite system into two semi-infinite parts, each with a surface (in red) parallel to the impurity plane. The impurity and the red surfaces are separated by a unit-cell spacing, which in this case is equal to $2 d_0$ for Bernal-stacked graphite and to $3 d_0$ for rhombohedral-stacked graphite.}
	\label{fig:plane_impurity}
\end{figure}

\section{Spectral function for semi-infinite graphite}\label{sec:SpectralFunctionSemiInfGraphite}

Numerical tight-binding is valid for a finite-width graphite slab, however it cannot be applied to semi-infinite systems. To calculate the surface states for a semi-infinite system we use the technique described in Ref.~[\onlinecite{Pinon2020a}].
In order to emulate a surface we add a plane-like impurity to the bulk system, as depicted in black in Fig.~\ref{fig:plane_impurity}, such that the plane is perpendicular to the $z$ axis. We assume that the impurity is described by $V = U \delta(z) \mathbb{I}$, where the identity matrix $\mathbb{I}$ has the same dimensions as $\mathcal{H}$. The impurity generates impurity-bound states on either side of it (i.e., on the surfaces highlighted in red in Fig.~\ref{fig:plane_impurity}). For large values of the impurity potential amplitude $U$, these impurity-induced states become the surface states of graphite. Throughout this work we will thus be using $U=100\,000$.

Mathematically, this can be translated into a calculation of the system Green's function in the presence of the impurity, which can be done exactly via T-matrix \cite{Byers1993,Salkola1996,Ziegler1996,Mahan2000,Balatsky2006,Bena2016}:
\begin{align}
G(\bs{k}_1, \bs{k}_2, i \omega_n) = 2\pi G_0(\bs{k}_1, i \omega_n) \delta\left( \bs{k}_1 - \bs{k}_2 \right) + \nonumber \\
G_0(\bs{k}_1, i \omega_n) T(\bs{k}_1, \bs{k}_2, i \omega_n) G_0(\bs{k}_2, i \omega_n)
\label{eq:PerturbedGreen}
\end{align}
where the T-matrix $T(\bs{k}_1, \bs{k}_2, i \omega_n)$ embodies the effect of all-order impurity-scattering processes, and is given by
\begin{align}
T(k_{1x}, k_{1y}, k_{2x}, k_{2y}, i \omega_n) = \delta_{k_{1x}, k_{2x}} \delta_{k_{1y}, k_{2y}} \times \nonumber \\
\left[ \mathbb{I} - U \int \frac{dk_z}{L_z} G_0(k_{1x}, k_{1y}, k_z, i \omega_n) \right]^{-1} U.
\label{eq:Tmatrix}
\end{align}
Here $L_z$ is the length of the Brillouin zone in the $k_z$ direction -- the direction perpendicular to the impurity plane 
\footnote{The integral over $k_z$, as well as other integrals mentioned below, are often approximated in our calculations by a Riemann sum or a trapezoid rule.}.
The zero-temperature retarded Green's function $\mathcal{G}(\bs{k}_1, \bs{k}_2, E)$ is obtained by an analytic continuation $i \omega_n \rightarrow E + i \delta$ with $\delta \rightarrow 0^+$. 

In what follows we assume that there is always some nonzero broadening of the energy $\delta$ accounting for inelastic scattering processes due to random ubiquitous disorder in the system and/or nonzero temperature. This allows us to make a qualitative comparison with the experiments where some nonzero widening of the energy is always observed. A qualitative comparison of the broadening parameter in graphite versus that of its two-dimensional surface can be found in Appendix \ref{App:Broadening}.

The surface Green's functions for a semi-infinite system correspond in this configuration to the Green's function on the planes shown in red in Fig.~\ref{fig:plane_impurity}. These can be obtained by performing a Fourier transform of $\mathcal{G}$ in the $z$ direction, and fixing $z$ at the appropriate value $z = \pm z_0$, with $z_0 = 2 d_0$ for ABA graphite and $z_0 = 3 d_0$ for ABC graphite:
\begin{align}
&\mathcal{G}_s(k_x, k_y, z = \pm z_0, E) = \nonumber \\
&\int \negthickspace \int \frac{dk_{1z}}{L_z} \frac{dk_{2z}}{L_z} \mathcal{G}(k_x, k_x; k_y, k_y; k_{1z}, k_{2z}; E) e^{i k_{1z} z} e^{-i k_{2z} z}.
\label{eq:surfaceGreen}
\end{align}
The corresponding surface spectral function is given by:
\begin{align}
A_s(k_x, k_y, E)\Big|_{z = \pm z_0} = - \frac{1}{\pi} \im \tr \mathcal{G}_s(k_x, k_y, z = \pm z_0, E) 
\label{eq:surfaceSpectral}
\end{align}

\begin{figure}[h]
	\centering
	\hspace*{-1.3cm}
	\includegraphics[width=0.61\linewidth]{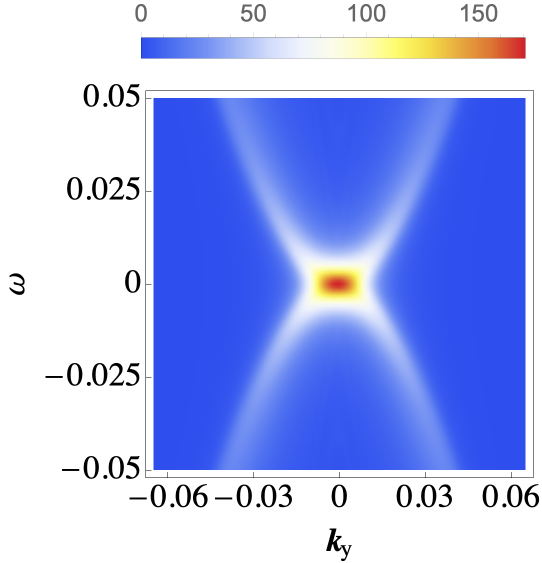}
	\caption{Surface spectral function of Bernal-stacked graphite around the $K$ point, at $z = 2 d_0$. We have taken $\gamma_3=0$ and $\delta=0.005$.}
	\label{fig:graphiteBernalKpointSpectral}
\end{figure}

In Fig.~\ref{fig:graphiteBernalKpointSpectral} we consider Bernal-stacked graphite and plot $A_s$ for $z = 2 d_0$ over the same range of energies and momenta as in Fig.~\ref{fig:graphiteBernalKpoint}. We recover a parabolic-shape band, as expected from previous knowledge about graphite. This is consistent with the tight-binding spectrum presented in Fig.~\ref{fig:graphiteBernalKpoint} which contains both the surface and the bulk bands for a finite graphite slab.

\begin{figure}[h!]
	\centering
	\includegraphics[width=0.45\linewidth]{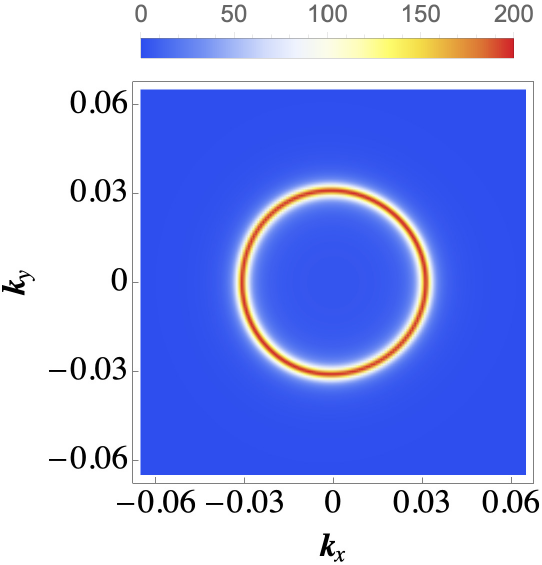}
	\includegraphics[width=0.45\linewidth]{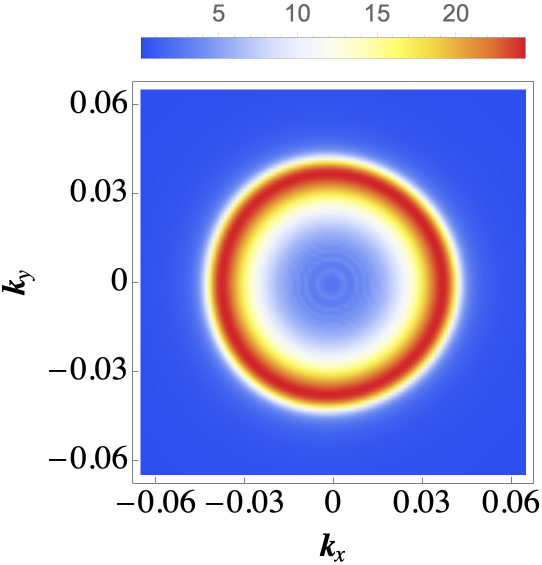}\\
	\includegraphics[width=0.45\linewidth]{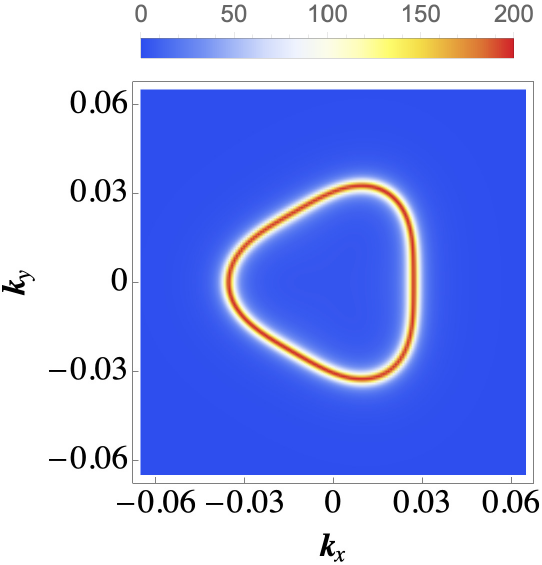}
	\includegraphics[width=0.45\linewidth]{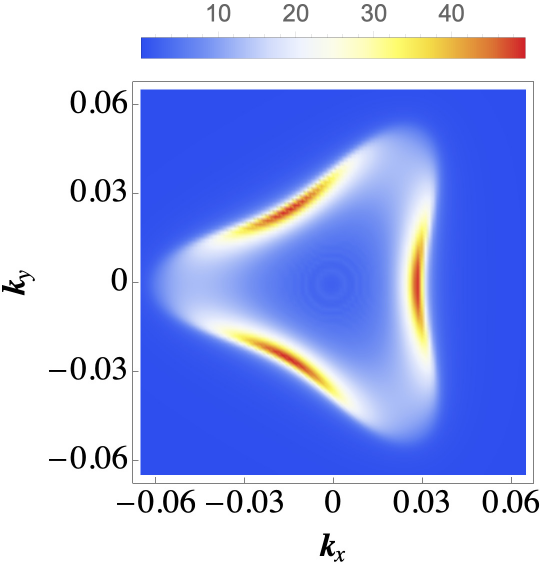}
	\caption{Surface spectral functions of bilayer graphene (left column) and Bernal-stacked graphite (right column), around the $K$ point calculated at $E=0.05$, without (top row, $\gamma_3=0$) and with trigonal warping (bottom row, $\gamma_3=-0.3$).  We have taken $\delta=0.005$.}
	\label{fig:BLGBernalwwoTW}
\end{figure}

In Fig.~\ref{fig:BLGBernalwwoTW} we plot the surface spectral functions at a given energy $E=0.05$ as a function of $k_x$ and $k_y$, both for bilayer graphene and Bernal-stacked graphite (left and right columns of Fig.~\ref{fig:BLGBernalwwoTW}, correspondingly). To provide better understanding, we compute the surface spectral functions with and without trigonal warping (top and bottom rows of Fig.~\ref{fig:BLGBernalwwoTW}, respectively). In the absence of the warping term the surface spectral function taken at a fixed energy seems to be rotationally symmetric in the $(k_x, k_y)$ plane, both for graphite and bilayer graphene. Once the trigonal warping term is introduced into the model, the results acquire a three-fold symmetry. The effect of this trigonal warping seems to be more pronounced for the surface states of graphite than for the BLG. This can be understood as follows: the warping is a result of inter-layer coupling terms, and hence the more layers the material contains the more pronounced the warping is expected to be. Moreover, we note that the surface spectral function of graphite is much less sharp, with equal energy lines much wider and less well defined than for BLG, as well as with a large residual intensity in the background. This suggests that quasiparticles have shorter lifetimes on the surface of graphite than in BLG, and thus a decrease in the coherence length for the surface states.

\begin{figure}[t!]
\centering
\includegraphics[width=0.61\linewidth]{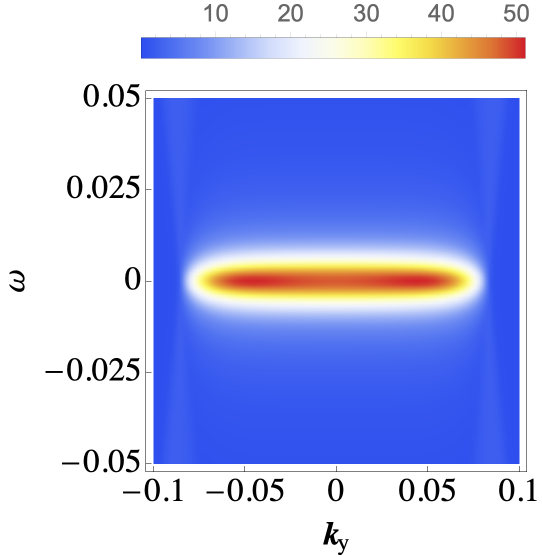}
\caption{Surface spectral function of rhombohedral-stacked graphite around the $K$ point, at $z = 3 d_0$. We have taken $\gamma_3=0$ and $\delta=0.005$. }
\label{fig:graphiteRhomboKpointSpectral}
\end{figure}

We have performed a similar analysis for rhombohedral graphite: the dependence of the surface spectral function on the energy and momentum is depicted in Fig.~\ref{fig:graphiteRhomboKpointSpectral}. We observe once more the formation of flat bands, consistent with the results of the tight-binding analysis of the fin ite-size graphite slabs (Fig.~\ref{fig:graphiteRhomboKpoint}). The extension of the flat band observed here for the semi-infinite graphite seems close to the one observed in the tight-binding calculations for 100 layers (the right panel in Fig.~\ref{fig:graphiteRhomboKpoint}), confirming that it is converging to a finite value when the width of the slab is going to infinity.

\begin{figure}[h]
	\centering
	\includegraphics[width=0.45\linewidth]{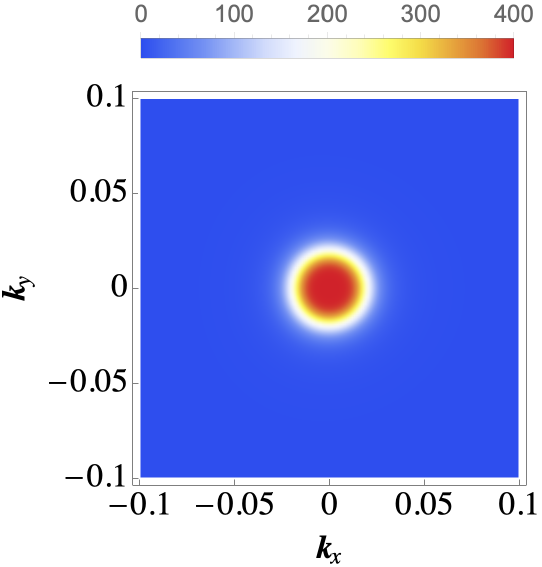}
	\includegraphics[width=0.45\linewidth]{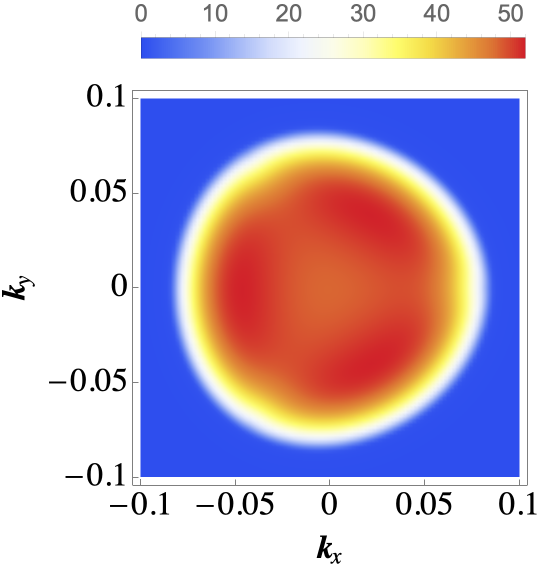}\\
		\includegraphics[width=0.45\linewidth]{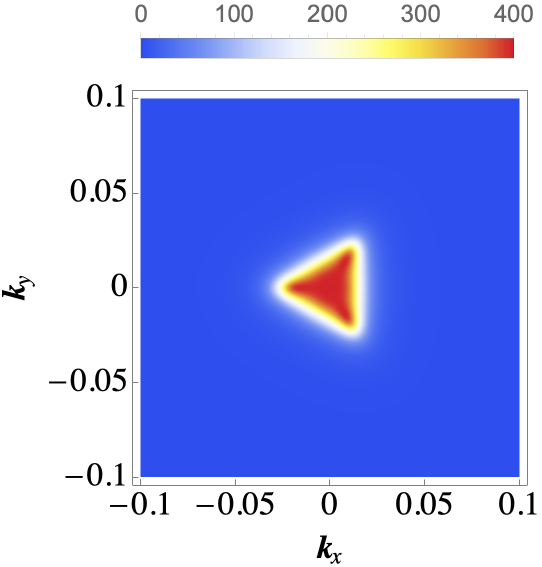}
	\includegraphics[width=0.45\linewidth]{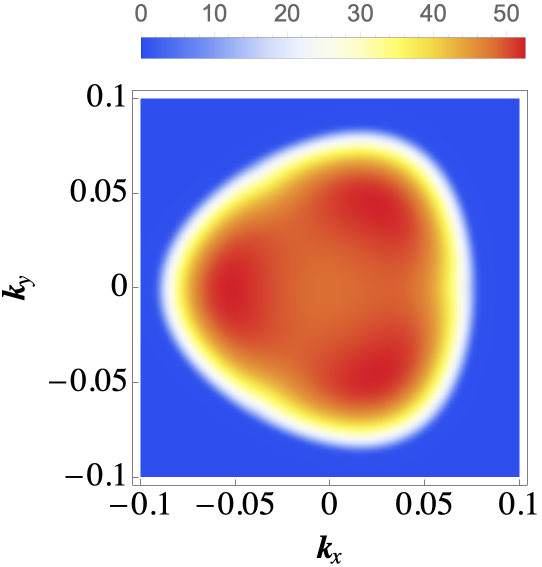}\\
		\includegraphics[width=0.45\linewidth]{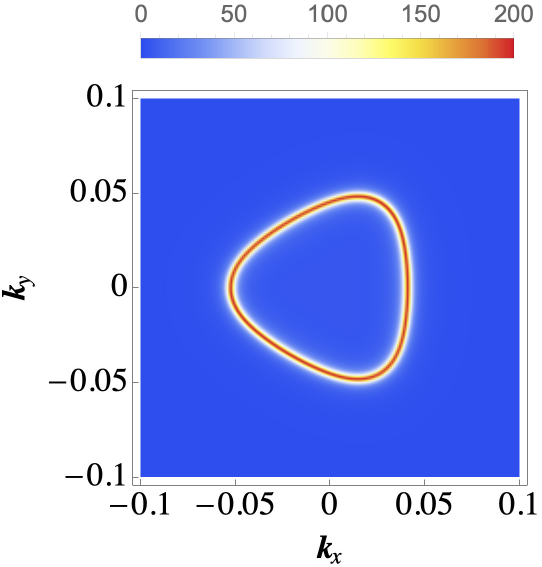}
	\includegraphics[width=0.45\linewidth]{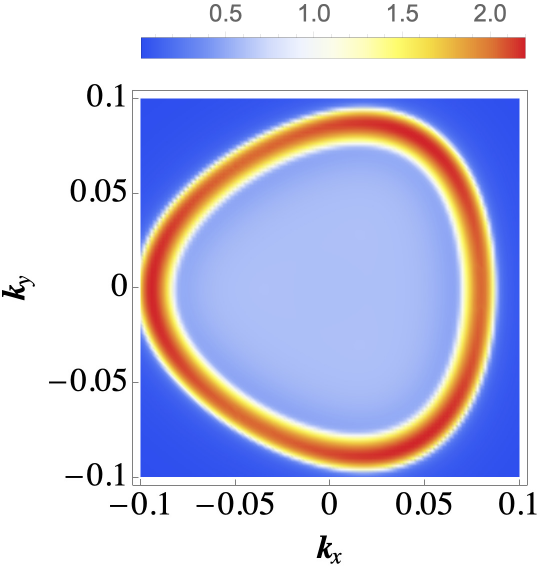}
	\caption{Spectral function of trilayer ABC graphene (left column) and surface spectral function for ABC graphite (right column), without trigonal warping $\gamma_3=0$ (top row) and with trigonal warping $\gamma_3=-0.3$ (bottom two rows) around the $K$ point at $E=0$ (top two rows) and $E=0.05$ (bottom row). We have taken $\delta=0.005$.}
	\label{fig:graphiteRhomboKpointSpectralkspace}
\end{figure}

Furthermore, in Fig.~\ref{fig:graphiteRhomboKpointSpectralkspace} we plot the surface spectral function in the $(k_x,k_y)$ plane, with and without trigonal warping. For comparison, in the left column we plot the spectral function of the ABC trilayer graphene. In this case, we can see that a very weak three-fold symmetry is present for graphite even when considering $\gamma_3 = 0$.  A much stronger one is observed on the other hand when adding the trigonal warping terms.




\section{Modification of the local density of states in the presence of an impurity}\label{sec:QPIpatterns}

In what follows, we calculate the modification of the LDOS for the systems described above in the presence of a localized impurity. For the surface of ABA and ABC graphite we start from the surface Green's functions described in Eq.~(\ref{eq:surfaceGreen}) and we apply the T-matrix formalism, which yields for the Fourier transform of the impurity-induced modifications of the density of states:
\begin{equation}
\delta\rho(\bs{k},E)= \frac{i}{2 \pi} \negthickspace\int \negthickspace\frac{d\bs{q}}{(2\pi)^2} \tr_{12}\left[g(\bs{q},\bs{k}, E)\right]
\label{eq:QPIintegral}
\end{equation}
where $d\bs{q} \equiv dq_x dq_z$,
\begin{align}
g(\bs{q},\bs{k}, E) \equiv \mathcal{G}_s (\bs{q},E) T(E) \mathcal{G}_s(\bs{q-k},E) - \nonumber \\
\mathcal{G}^*_s(\bs{q-k},E) T^*(E) \mathcal{G}^*_s (\bs{q},E),
\label{eq:QPIgfunction}
\end{align}
and the $T$-matrix can be found via
\begin{equation}
T(E) = \left[\mathbb{I} - V \negthickspace\int\negthickspace \frac{d\bs{q}}{(2\pi)^2}\, \mathcal{G}_s(\bs{q},E) \right]^{-1} V.
\label{eq:QPITmat}
\end{equation}
The impurity matrix $V$ has the same dimensions as the Green's function matrix ($4 \times 4$ for BLG and ABA graphite, and $6 \times 6$ for TLG and ABC graphite), and is a diagonal matrix with a nonzero element at the position corresponding to the layer-sublattice combination at which the impurity is localized (e.g. $V_{11}/V_{22} \neq 0$ for an atom in the top layer on the A/B sublattice, and so on). Note that we take the trace $\tr_{12}$ in Eq.~(\ref{eq:QPIintegral}) only over the first two components of the matrix corresponding to the two top-layer atoms, since this is the contribution to the density of states that is usually measured experimentally. 

For bilayer graphene and trilayer graphene the same formalism applies, but with $\mathcal{G}_s$ in Eq.~(\ref{eq:QPIgfunction}) and (\ref{eq:QPITmat}) replaced by the unperturbed Green's function calculated via Eq.~(\ref{eq:MatsubaraGreen}) starting from the unperturbed BLG and TLG Hamiltonians in Eqs.~(\ref{hh1},\ref{hh2}).

We focus on the modification of the LDOS both in momentum space and in real space. The modulation of the LDOS due to the impurity in the T-matrix formalism in real space is given by:
\begin{equation}
\delta \rho(\bs{r},E) = -\frac{1}{\pi} \im \tr_{12} \left[\mathcal{G}_s(\bs{r},E) T(E)\mathcal{G}_s(-\bs{r},E) \right],
\end{equation}
where $\mathcal{G}_s(\bs{r},E) = \int \frac{d\bs{q}}{(2\pi)^2} \mathcal{G}_s(\bs{q},E) e^{i \bs{q r}} $ is the Green's function in real space calculated via a Fourier transform.

\subsection{Fourier transform of the LDOS}

\begin{figure}[h]
	\centering
	\hspace*{-0.6cm}
	\includegraphics[width=0.45\linewidth]{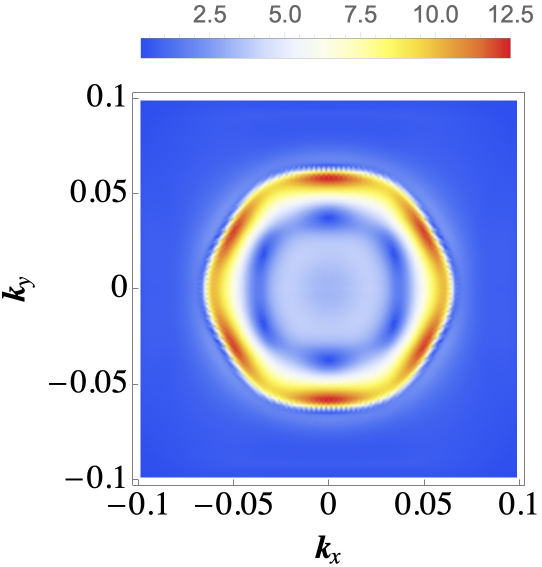}
	\includegraphics[width=0.45\linewidth]{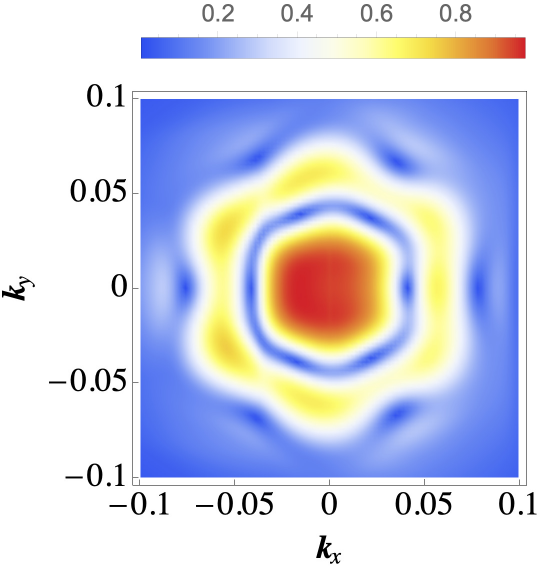}
	\caption{Quasiparticle interference patterns for bilayer graphene (left panel) and Bernal-stacked graphite (right panel) around the $K$ point calculated at $E = 0.05$. We consider a trigonal warping value $\gamma_3 = -0.3$, and we take $V = -10, \gamma_0 = - 3.3$.}
	\label{fig:QPIBernal}
\end{figure}

The results for the quasiparticle interference patterns for ABA graphite and BLG are presented in Fig.~\ref{fig:QPIBernal}. For BLG we observe a high-intensity hexagonal contour in agreement with the recent high-resolution experimental results \cite{Joucken2021} and in contrast to the circular contour observed in BLG without trigonal warping \cite{Bena2008}.
For ABA graphite a hexagonal feature is also observed. The structure of these features may be inferred from the corresponding surface spectral function shown in Fig.~\ref{fig:BLGBernalwwoTW}. While the surface spectral function for bilayer graphene consists of a sharp uniform triangle, the spectral function for the surface of ABA graphite consists of a blurred deformed triangle with well-pronounced maxima on the sides. The curvature of the triangle sides, and the intensity higher on the sides than in the corners, imply an enhancement of the scattering inside a single side, consistent with a high-intensity feature at the center of the QPI hexagone. This feature is absent from the QPI BLG pictures. Moreover, the blurring of the spectral function feature implies a wider distribution of the allowed scattering states and thus a wider and more blurred QPI feature.  

\begin{figure}[h]
	\centering
	\hspace*{-0.6cm}
		\includegraphics[width=0.45\linewidth]{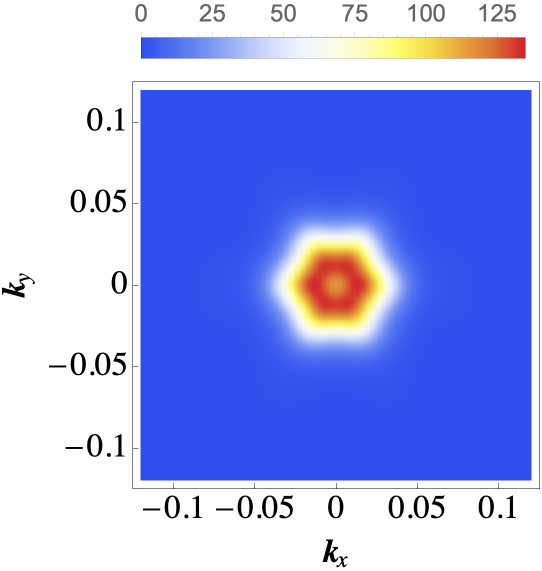}
	\includegraphics[width=0.45\linewidth]{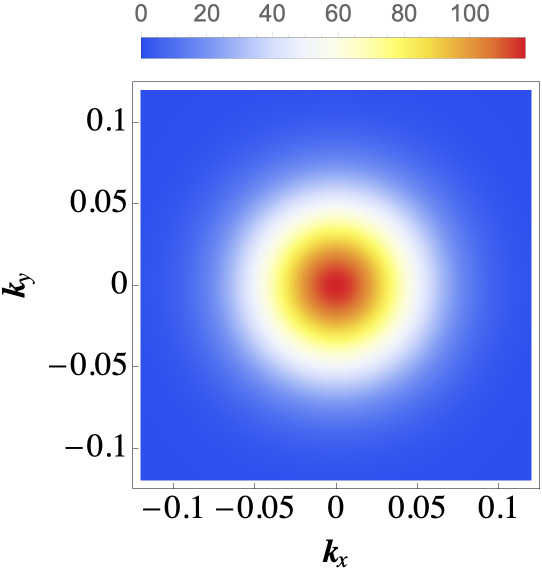}\\
	\hspace*{-0.6cm}
		\includegraphics[width=0.45\linewidth]{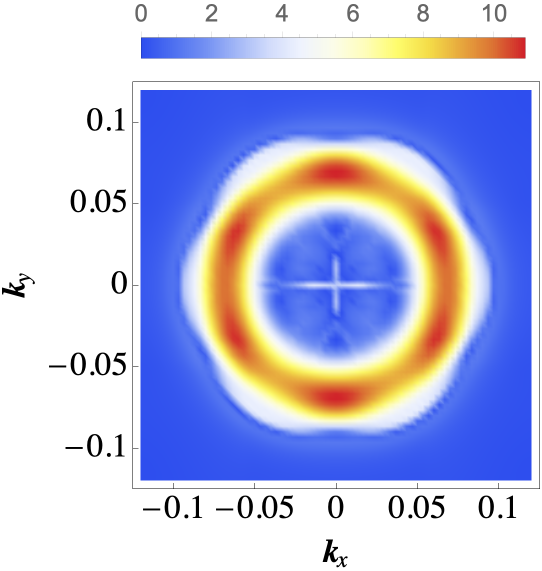}
	\includegraphics[width=0.45\linewidth]{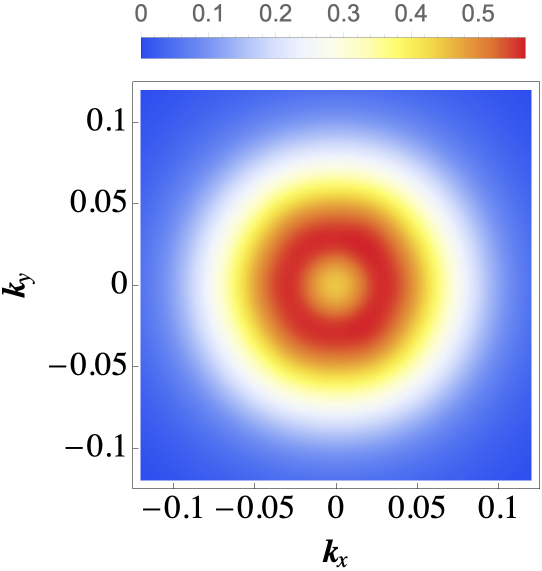}
	\caption{Quasiparticle interference features for ABC graphene (left) and rhombohedral-stacked graphite (right) around the $K$ point at $E=0$ (top row), and $E=0.05$ (bottom row). We take $\delta=0.005\,$, and $\gamma_{3}=-0.3$ and $V=-33$.}
	\label{fig:QPIrhombo}
\end{figure}

We also compute the quasiparticle interference features for ABC graphite and ABC trilayer graphene. The results are shown in Fig.~\ref{fig:QPIrhombo}. As expected from the surface spectral function calculated at $E = 0$, the graphite QPI features are blurred due to the fact that many states with equal spectral weight exist at the surface, both at zero-energy, inside the flat band, as well as at higher energy, as we can see in Fig.~\ref{fig:graphiteRhomboKpointSpectralkspace}. The six-fold symmetry arising due to the trigonal warping effects is harder to see in the QPI features for ABC graphite, though it is still visible in real space, as we will show in the next section. Note that in the absence of trigonal warping our findings are consistent with the analytical calculations of Fourier-transformed local density of states in rhombohedral $N$-layer graphene from Ref.~[\onlinecite{Dutreix2016}]. 

We note that all the QPI features above simulate an average response of impurities being localised not on one of the sublattices, but equally distributed on all the sublattices. In order to generate the QPI results for such a configuration we have used an impurity matrix described as an identity matrix in the full matrix sublattice space, which is equivalent to having equal impurity contributions from all sublattices.

\subsection{Real-space profile of impurity states}

\begin{figure}[h]
	\centering
	\hspace*{-0.6cm}
	\includegraphics[width=0.42\linewidth]{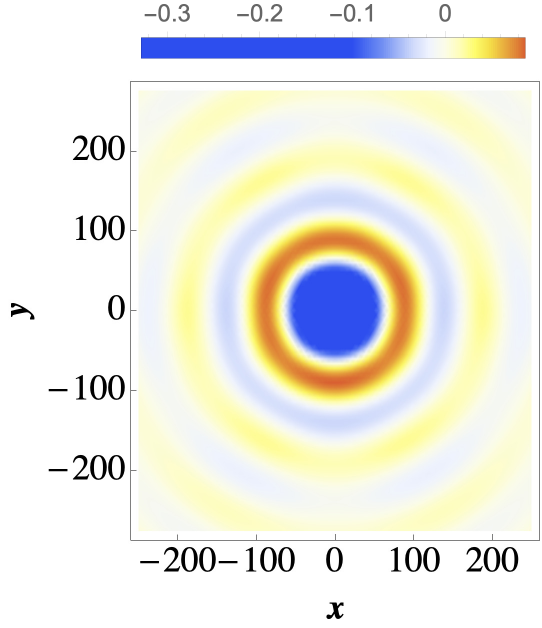}
	\includegraphics[width=0.42\linewidth]{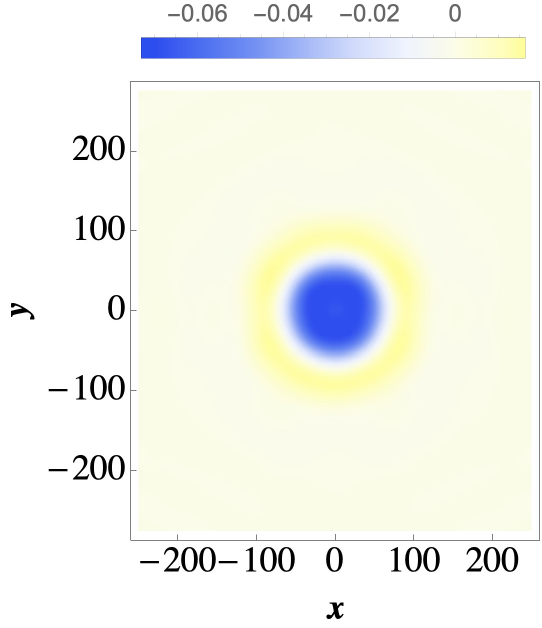}\\
	\hspace*{-0.6cm}
	\includegraphics[width=0.42\linewidth]{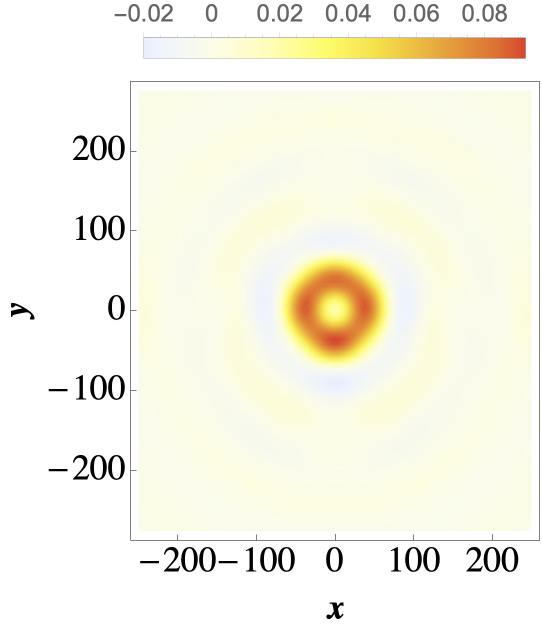}
	\includegraphics[width=0.42\linewidth]{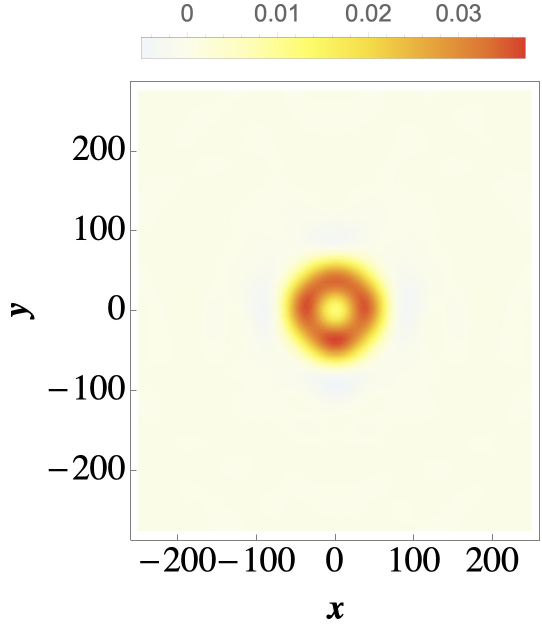}\\
	\hspace*{-0.6cm}
	\includegraphics[width=0.42\linewidth]{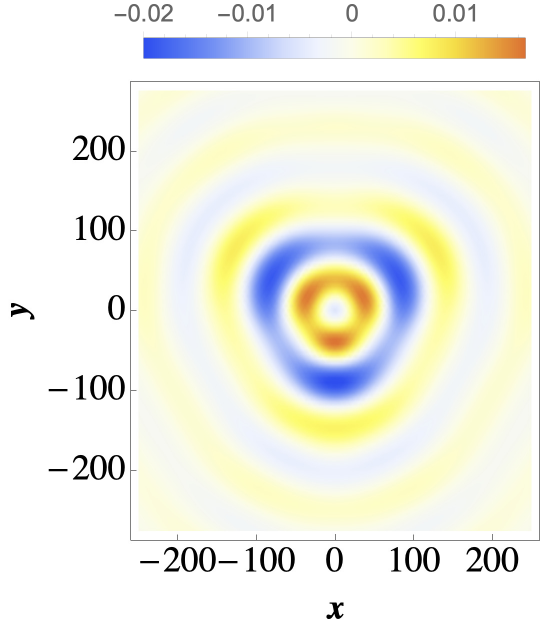}
	\includegraphics[width=0.42\linewidth]{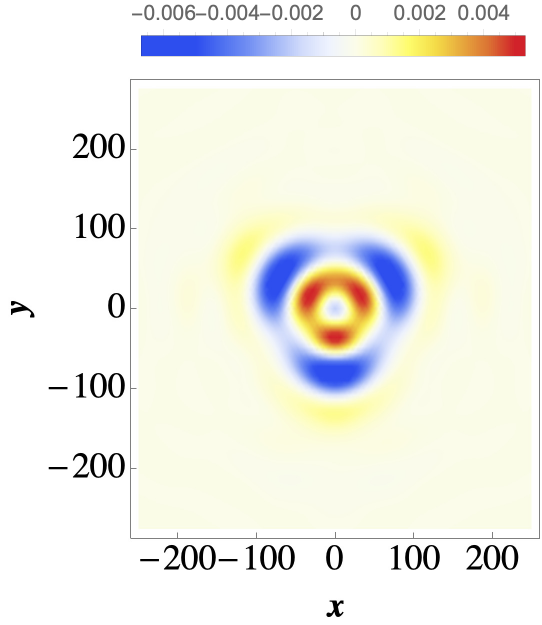}\\
	\hspace*{-0.6cm}
	\includegraphics[width=0.42\linewidth]{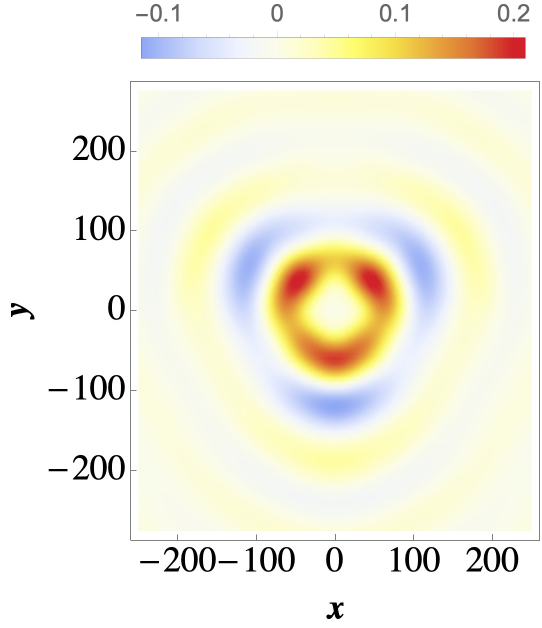}
	\includegraphics[width=0.42\linewidth]{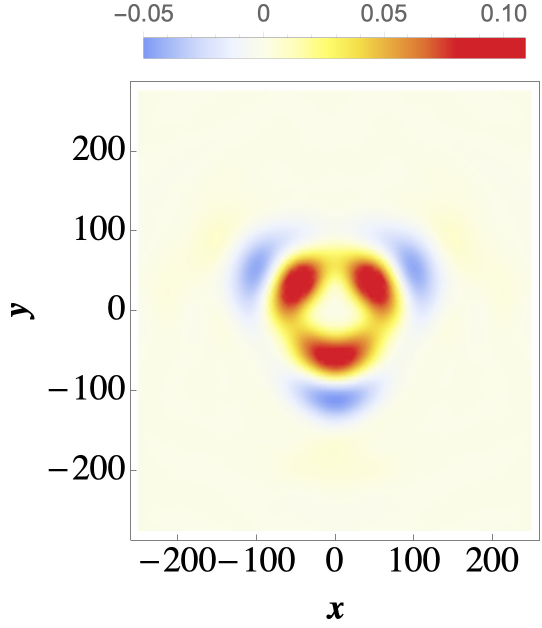}
	\caption{Impurity-induced corrections to the local density of states in bilayer graphene (left column row) and Bernal-stacked graphite (right column) calculated at $E=0.05$. We set $V=-33$ and $\gamma_3=-0.3$. The four rows correspond to impurities located on sites A1/B1/A2/B2. The LDOS corrections are given in units of the value of the unperturbed background. The results for bilayer graphene are in a good agreement with the analytical calculations presented in Ref.~[\onlinecite{Kaladzhyan2021}].}
	\label{fig:QPIBernalRS}
\end{figure}

We now focus on the fluctuations of the LDOS in real space in the vicinity of different types of impurities. Thus, in Fig.~\ref{fig:QPIBernalRS} we plot the correction to the local density of states both for bilayer graphene and Bernal-stacked graphite. We consider four different types of impurities localized on the A/B atoms of the top/bottom layer of the unit cell. First of all, we observe that the LDOS profile of the impurity states depends strongly on which sites the impurity is located. As also noted in Ref.~[\onlinecite{Joucken2021}] certain impurities give rise to almost rotationally symmetric patterns which do not reflect the presence of trigonal warping, while other impurities clearly reflect the three-fold symmetry originating from trigonal warping terms.
\begin{figure}[h]
	\centering
	\hspace*{-0.6cm}
	\includegraphics[width=0.42\linewidth]{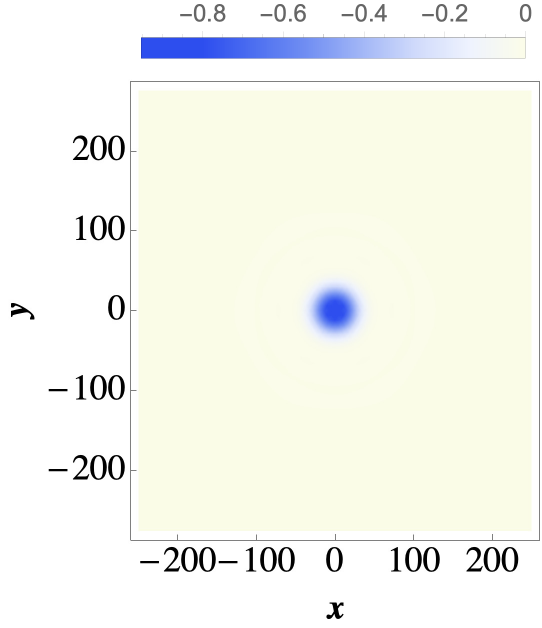}
	\includegraphics[width=0.42\linewidth]{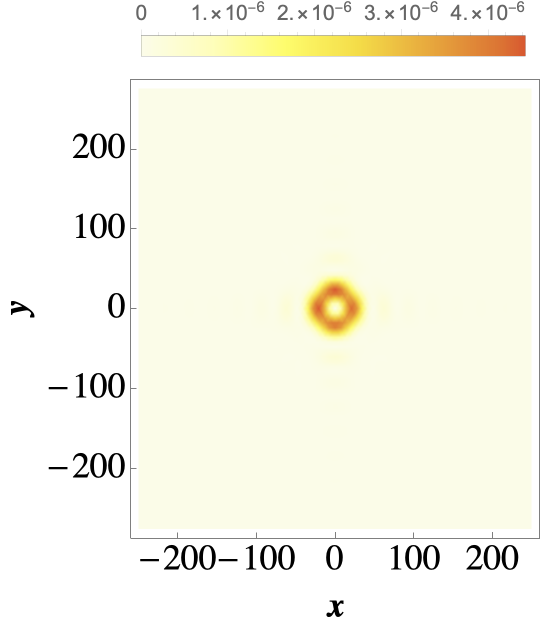}\\
	\hspace*{-0.6cm}
	\includegraphics[width=0.42\linewidth]{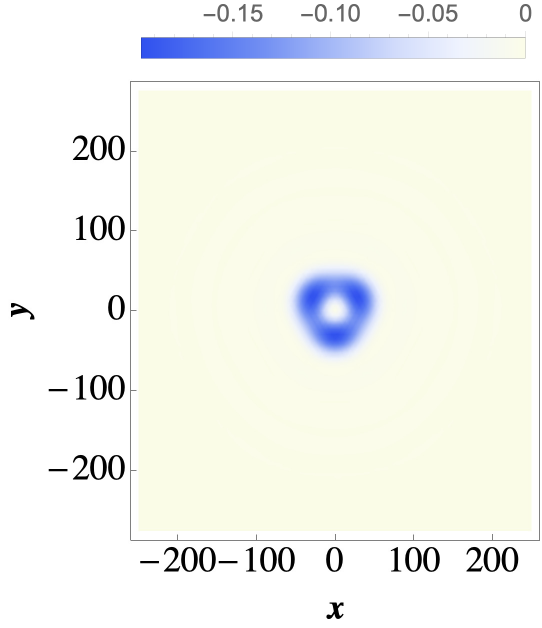}
	\includegraphics[width=0.42\linewidth]{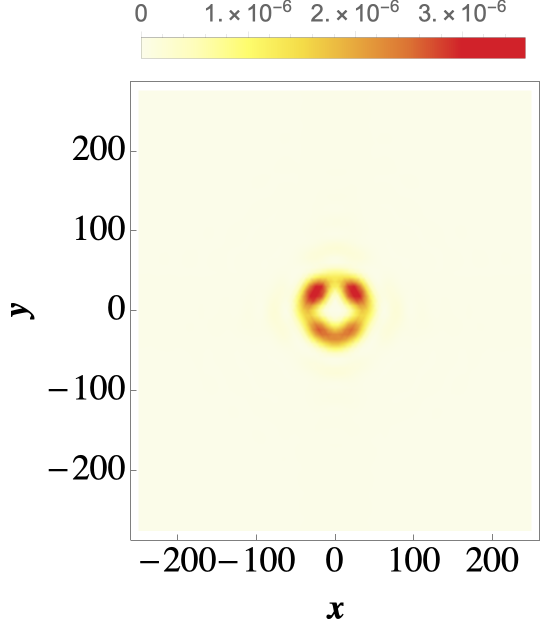}\\
	\hspace*{-0.6cm}
	\includegraphics[width=0.42\linewidth]{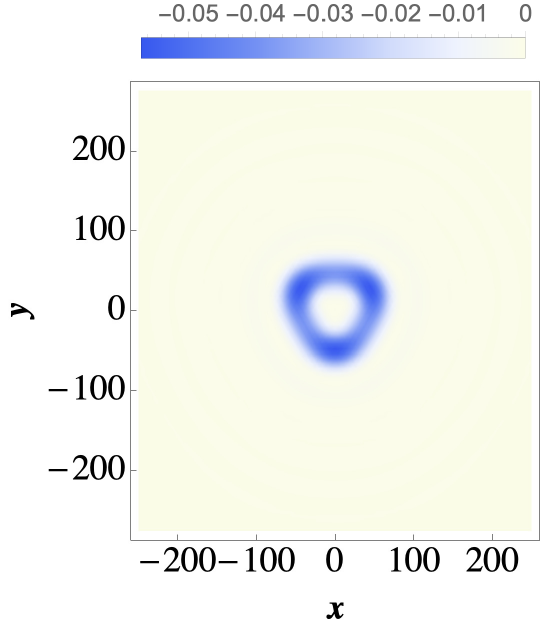}
	\includegraphics[width=0.42\linewidth]{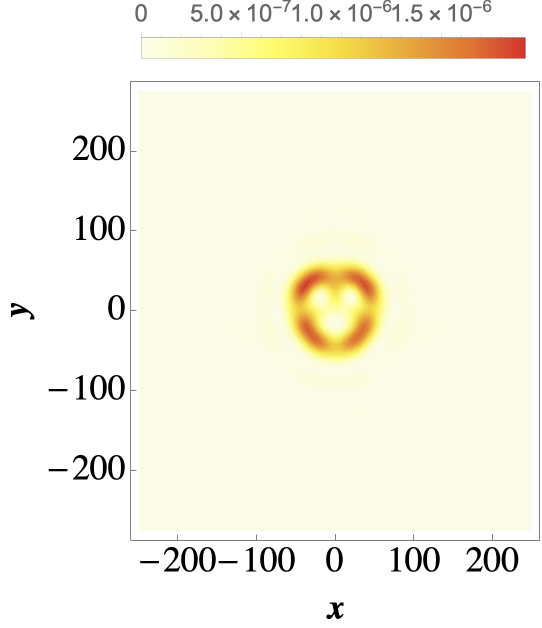}
	\caption{Impurity-induced corrections to the local density of states in ABC graphite at $E=0$, $\gamma_{3}=-0.3$, $V=-33$ and $\delta=0.005$. The panels correspond to impurities located on sites A1/B1/A2/B2/A3/B3. The LDOS corrections are given in units of the value of the unperturbed background. }
	\label{fig:LDOStrilayergraphene}
\end{figure}
The most important observation is that the localization length of the impurity-induced states is much smaller for graphite than for bilayer graphene. This is consistent with the reduction of the quasiparticle lifetime, which corresponds to the blurring of the spectral function of the surface states, and can be intuitively understood as being the result of the three-dimensional character of graphene for which quasiparticles have an extra spatial dimension to be scattered in. In contrast, in bilayer graphene the oscillations in the local density of states stay very well-pronounced much further away from the impurity. Such a difference in localization length seems to be in general governed by the dimension of the problem: the localization length is shorter in higher dimensions, as it was observed for instance for Yu-Shiba-Rusinov states in three- versus two-dimensional superconductors \cite{Yazdani1997,Shuai-Hua2008,Menard2015}. The evolution from two layers to a large number of layers is explored in more details in Sec.~VII.

Note that the three-fold character of the observed features is much more pronounced when the impurity is located in the bottom layer. We believe that this stems from the fact that we only plot the intensity of the LDOS in the top layer. Thus, when an impurity is located in the top layer we expect the features generated in the same layer to be roughly similar to those generated in monolayer graphene, i.e., circularly symmetric at long distances, with the trigonal warping not manifesting strongly since the main visible effects are the intra-layer ones. On the other hand, when the impurity is in the bottom layer, the effects visible in the top layer will be mostly due to the inter-layer hopping terms among which the trigonal warping term plays an important role, and thus the symmetry of the impurity feature should show a strong enhanced three-fold character.

For ABC graphite, in Fig.~\ref{fig:LDOStrilayergraphene} we compute the impurity-induced corrections to the local density of states for six different types of impurities, localized on the A/B sublattices and in one of the three layers of the ABC graphite unit cell.  Same as for the ABA graphite we see that some of the features are fully circular, while others reflect, as expected, the three-fold symmetry originating from trigonal warping. Also, same as for ABA graphite, the oscillations decay very fast with the distance from the impurity.

\section{Experimental data}\label{sec:ExperimentalData}

We performed STM measurements on graphite and multilayer graphene ($\sim10$ layers) to verify our theoretical findings. We first describe the experimental details. The STM measurements were conducted in ultra-high vacuum with pressures better than $1 \times 10^{-10}\,$mbar at $4.8\,$K in a Createc LT-STM. The bias is applied to the sample with respect to the tip. The tips were electrochemically-etched tungsten tips, which were calibrated against the Shockley surface state of Au(111) prior to measurements. The graphite (“Flaggy Flakes” and “Graphenium Flakes” from NGS Naturgraphit) sample was exfoliated \textit{in situ} and introduced in the STM head within seconds after the exfoliation. The graphene multilayer heterostructures were stacked on hBN using a standard polymer-based transfer method \cite{Zomer2011}. A graphene flake exfoliated on a methyl methacrylate (MMA) substrate was mechanically placed on top of a $\sim50\,$nm thick hBN flake that rests on a SiO$_2$/Si++ substrate where the oxide is $285\,$nm thick. Subsequent solvent baths dissolve the MMA scaffold. After the Graphene/hBN heterostructure is assembled, an electrical contact to graphene is made by thermally evaporating $7\,$nm of Cr and $200\,$nm of Au using a metallic stencil mask. The single-terminal device is then annealed in forming gas (Ar/H$_2$) for six hours at $400\,^{\circ}$C to reduce the amount of residual polymer left after the graphene transfer. To further clean the surface of the sample, the heterostructure is mechanically cleaned using an AFM \cite{Goossens2012,Ge2020}. Finally, the heterostructure is annealed under UHV at $400\,^{\circ}$C for seven hours before being introduced into the STM chamber.

\begin{figure}
	\centering
	\includegraphics[width=0.98\linewidth]{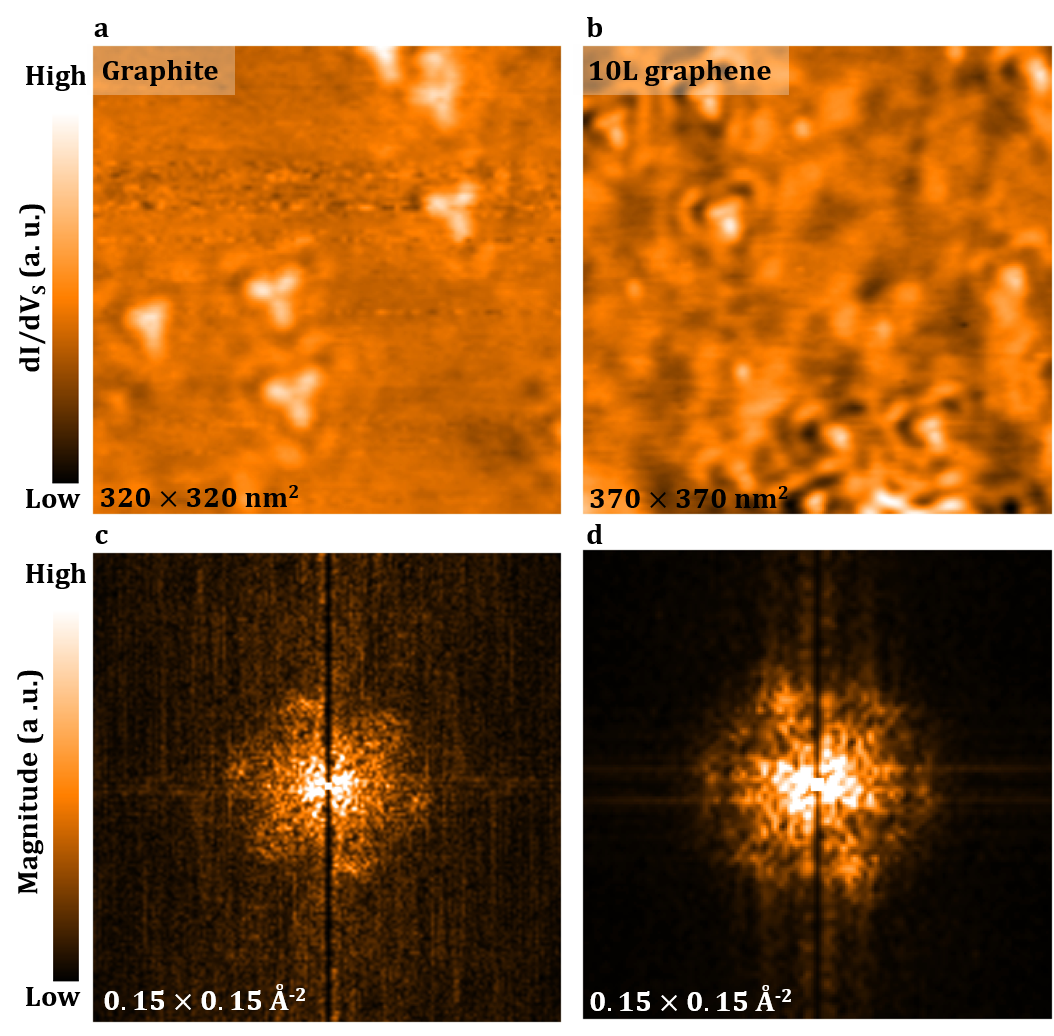}
	\caption{Experimental measurements of the real-space and Fourier-space quasiparticle interference patterns for impurities in graphite and many-layer graphene. Panels (a) and (b): $dI/dV_S$ maps obtained on, respectively, graphite and ten-layer-thick graphene film supported by hexagonal boron nitride. Panels (c) and (d): zoom-ins around the origin of the respective fast Fourier transforms. The $dI/dV$ maps were measured at a tip-sample bias $V_S = +5\,$mV, while applying a $V_{\mathrm{AC}}=3.5\,$mV excitation with a lock-in amplifier, and scanning at a constant current $I=0.5\,$nA and $I=0.25\,$nA for panels (a) and (b), respectively.}
	\label{fig:ExpQPIgraphitegraphene}
\end{figure}

We show in Figs.~\ref{fig:ExpQPIgraphitegraphene}a and \ref{fig:ExpQPIgraphitegraphene}b typical $dI/dV_S$ spatial maps of graphite and $\sim10$-layer graphene film, respectively. These maps were acquired at low tip-sample bias (around $5\,$mV), so that they are essentially proportional to the LDOS at the Fermi level. As for the simulations presented above (Fig.~\ref{fig:QPIBernalRS}), one can see the real-space scattering patterns around defects are localized around the scattering centers, contrary to what was observed for bilayer graphene at high charge carriers concentration \cite{Joucken2021}, where much longer localization lengths were observed. Also, the scattering pattern around the defect displays a salient three-fold symmetry, originating from the trigonal warping, as discussed above. The FFT signatures for $dI/dV_S$ spatial maps in Figs.~\ref{fig:ExpQPIgraphitegraphene}a and \ref{fig:ExpQPIgraphitegraphene}b are presented in Figs.~\ref{fig:ExpQPIgraphitegraphene}c and \ref{fig:ExpQPIgraphitegraphene}d, correspondingly. The latter also agree well with the calculations presented above (the right panel in Fig.~\ref{fig:QPIBernal}), with high intensity at the center of the scattering pattern, corresponding to the short localization length visible in the real space images.


\section{Dependence on the number of layers and energy}\label{sec:EvolutionWithLayersNumber}
To understand the experimental observations, which show a majority of features with a strong triangular symmetry, we first point out that, as noted also in Ref.~[\onlinecite{Joucken2021}], impurities fully localized on the topmost layer give rise to circularly symmetric features. On the other hand, experimentally most of the features observed in graphite are trigonally warped. We believe that one origin of this discrepancy, is, as also explained in Ref.~[\onlinecite{Joucken2021}], the fact that the impurities are not exactly fully localized in the top layer, but also have some component in the bottom layer. This components gives rise to a very strongly warped component to the LDOS in the top layer, and thus may explain the majority of the trigonally warped features in graphite. 

Moreover, we present here two more arguments which show that impurities in graphite are susceptible to show a very pronounced trigonal warping: 1. the closest we are to the Dirac point, and 2. the larger the number of the layers in the sample, the more pronounced the trigonal warping of the features.

\begin{figure}[t]
	\centering
	\hspace*{-0.6cm}
	\includegraphics[width=0.42\linewidth]{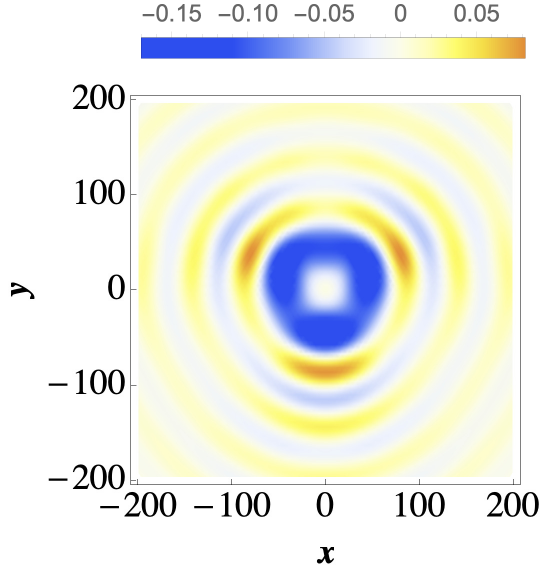}
	\includegraphics[width=0.42\linewidth]{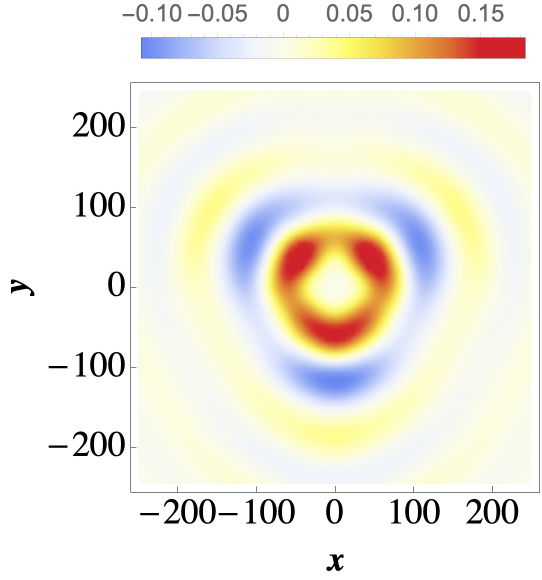}\\
	\hspace*{-0.6cm}
	\includegraphics[width=0.42\linewidth]{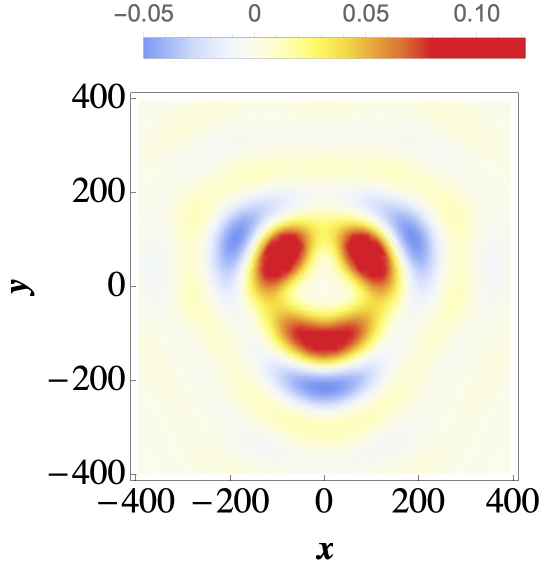}
	\includegraphics[width=0.42\linewidth]{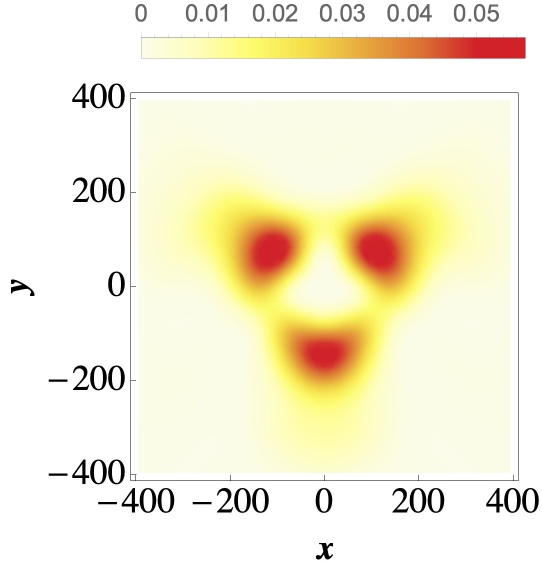}
	\caption{Impurity-induced corrections to the local density of states in bilayer graphene at four different energies, $E=0.25$ (top left), $0.05$ (top right), $0.02$ (bottom left), and $0$ (bottom right). We take $\gamma_{3}=-0.3$, $\delta=0.005$ and $V=-33$. The LDOS corrections are given in units of the value of the unperturbed background.}
	\label{fig:LDOSbilayerdiffE}
\end{figure}

For this, we study the dependence of the real-space patterns on the number of layers and on energy. Thus we consider systems with 2, 3, 4 and 8 layers, as well as a range of energies from $0$ to $250\,$meV. For all these different cases, we consider an ABA stacking and an impurity located on the B2 site. We note that for low energy, close to the Dirac point, the effects of the triangular warping are most pronouned, and the asymmetry of the triangular features becomes less visible when increasing the energy (see Fig.~\ref{fig:LDOSbilayerdiffE}). 

\begin{figure}[t]
	\centering
	\hspace*{-0.6cm}
	\includegraphics[width=0.42\linewidth]{imp-B2-blg-gt03-w05-dt0005-V33.pdf}
	\includegraphics[width=0.42\linewidth]{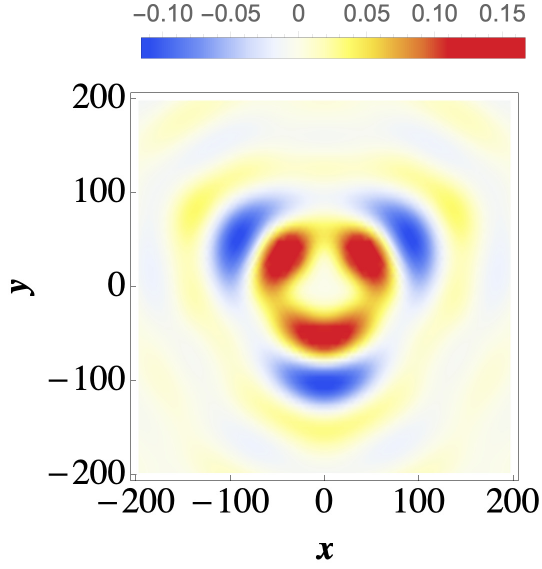}\\
	\hspace*{-0.6cm}
	\includegraphics[width=0.42\linewidth]{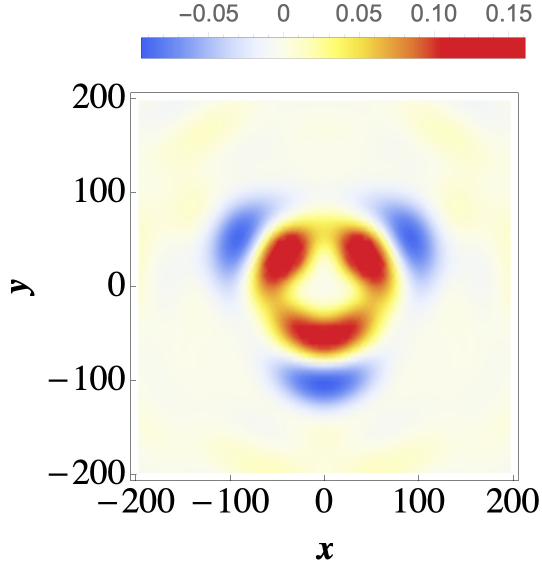}
	\includegraphics[width=0.42\linewidth]{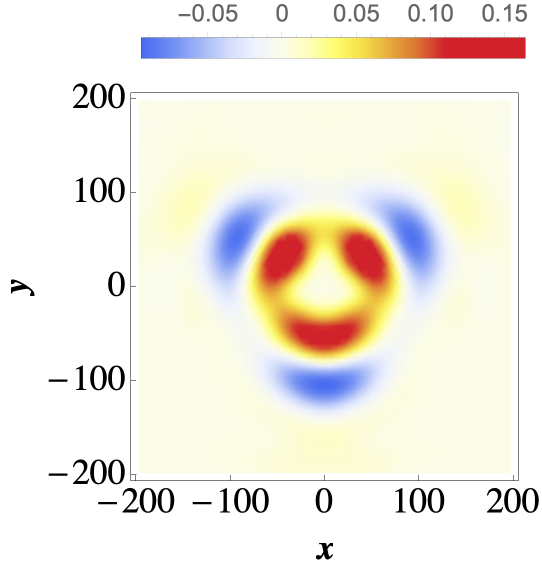}
	\caption{Impurity-induced corrections to the local density of states in bilayer graphene (top right) and in a 3-layer (top left), 4-layer (bottom left) and 8-layer (bottom right) systems at $E=0.05$. We take $\gamma_{3}=-0.3$, $\delta=0.005$ and $V=-33$. The LDOS corrections are given in units of the value of the unperturbed background.}
	\label{fig:LDOSmultilayergraphene}
\end{figure}

A similar observation can be made about changing the number of layers: the thicker samples show the strongest triangular features (see Fig.~\ref{fig:LDOSmultilayergraphene}). The details of the technique used to obtained the real-space impurity features in multilayer systems are presented in App.~\ref{App:LDOSmultilayer}.  Note that our results on multilayer graphene confirm the validity of the technique used in the previous sections to study semi-infinite systems, the results for the 8-layer system in Fig.~\ref{fig:LDOSmultilayergraphene} are basically identical to those corresponding to the semi-infinite system (lower right panel of Fig.~\ref{fig:QPIBernalRS}). 

The last observation is that, when increasing the number of layers, the oscillations decay faster and faster, and indeed for the semi-infinite system results presented in the previous sections, the impurity features are strongly localized close to the impurity. We observe that something similar happens also when decreasing the energy. Thus the highest energy features are the most long-lived, as well as the most circular. This is consistent with the measurements presented in Fig. 2 of Ref.~[\onlinecite{Joucken2021}], which reveal that at low gate voltages the oscillations are harder to see, decaying very fast, and showing a mostly triangular character, while at higher gates they decay much slower and become rather circular.

\begin{figure}
	\centering
	\includegraphics[width=0.5\linewidth]{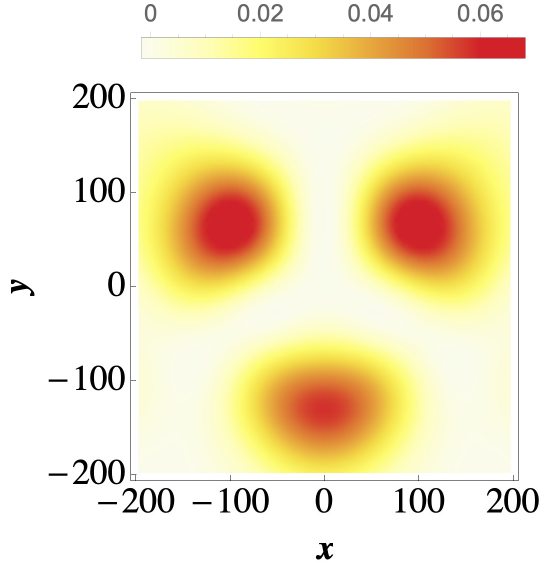}
	\caption{Impurity-induced corrections to the local density of states in an 8-layer system at $E=0$. We take $\gamma_{3}=-0.3$, $\delta=0.005$ and $V=-33$. The LDOS corrections are given in units of the value of the unperturbed background. }
	\label{fig:LDOS8layergraphene}
\end{figure}
Thus, by taking a very wide system such as the 8-layer graphene at zero energy (see Fig.~\ref{fig:LDOS8layergraphene}) we recover indeed the strongest triangular asymmetry.

These findings indicate that, consistent with the experimental observations, strongly trigonally warped features should be dominant in graphite at energies close to the Dirac point. Also, as also pointed out in Ref.~[\onlinecite{Joucken2021}], it appears that very localized impurities with a dominant component on a single atom in the top layer are not very common, since these would yield rather circular features: it seems that most of the impurities observed experimentally in graphene show some impurity potential component in the second layer, which introduces a strong triangular warping of the impurity features.

\section{Conclusions}\label{sec:Conclusion}

Using the technique developed in Ref.~[\onlinecite{Pinon2020a}] and [\onlinecite{Pinon2020b}], we have studied the surface spectral function, as well as the impurity-induced oscillations on the surface of Bernal-stacked and rhombohedral-stacked graphite, and compare the results to those obtained in bilayer and trilayer graphene. We have shown that this analytical technique is very well suited to the study of the surface physics of graphite; for example it allows to recover the flat-band surface states of rhombohedral-stacked graphite. Our first main observation was that the spectral function for graphite surfaces shows a decrease in coherence and quasiparticle lifetime compared to that of bilayer graphene, as well as stronger features associated with the trigonal warping. Secondly we have demonstrated that, independent of the type of stacking---Bernal or rhombohedral---surface impurity-induced oscillations decay much faster in graphite than in two- and three-layer graphene. Furthermore, these oscillations strongly reflect the presence of trigonal warping.  Most importantly, we have shown that our analytical results are in good agreement with the experimental data. 

\acknowledgments

J.V.J. acknowledges support from the National Science Foundation under award DMR-1753367 and the Army Research Office under contract W911NF-17-1-0473.

\bibliography{biblio_graphite}

\newpage
\widetext
\appendix

\section{Obtaining the LDOS correction for multilayer graphene}\label{App:LDOSmultilayer}

The process of obtaining the correction to the LDOS for 3, 4 or 8 multilayer graphene systems is the same as the one for bilayer graphene described in the main text, with the starting Hamiltonian $\mathcal{H}(\bs{k})$ being replaced by the multilayer Hamiltonian. If $N$ is the number of layers, then the Hamiltonian is a matrix of size $2N \times 2N$, corresponding to the number of layers and the two sublattices $A$ and $B$. If we define

\begin{align}
h_0(\bs{k}) &= - \gamma_0 \left[1 + 2 e^{- i \frac{3}{2} a_0 k_x} \cos \left(\frac{\sqrt{3}}{2} a_0 k_y \right) \right], \\
h_3(\bs{k}) &= - \gamma_3 \varepsilon,
\end{align}
the 8-layer Hamiltonian can be written as:

\setcounter{MaxMatrixCols}{20}
\begin{equation}
\nonumber \mathcal{H}(\bs{k}) = 
\begin{pmatrix}
0 & h_0(\bs{k}) & 0 & h_3(\bs{k}) & 0 & 0 & 0 & 0 & 0 & 0 & 0 & 0 & 0 & 0 & 0 & 0\\
h_0^*(\bs{k}) & 0 & \gamma_1 & 0 & 0 & 0 & 0 & 0 & 0 & 0 & 0 & 0 & 0 & 0 & 0 & 0\\
0 & \gamma_1 & 0 & h_0(\bs{k}) & 0 & \gamma_1 & 0 & 0 & 0 & 0 & 0 & 0 & 0 & 0 & 0 & 0\\
h_3^*(\bs{k}) & 0 & h_0^*(\bs{k}) & 0 & h_3^*(\bs{k}) & 0 & 0 & 0 & 0 & 0 & 0 & 0 & 0 & 0 & 0 & 0\\
0 & 0 & 0 & h_3(\bs{k}) & 0 & h_0(\bs{k}) & 0 & h_3(\bs{k}) & 0 & 0 & 0 & 0 & 0 & 0 & 0 & 0\\
0 & 0 & \gamma_1 & 0 & h_0^*(\bs{k}) & 0 & \gamma_1 & 0 & 0 & 0 & 0 & 0 & 0 & 0 & 0 & 0\\
0 & 0 & 0 & 0 & 0 & \gamma_1 & 0 & h_0(\bs{k}) & 0 & \gamma_1 & 0 & 0 & 0 & 0 & 0 & 0\\
0 & 0 & 0 & 0 & h_3^*(\bs{k}) & 0 & h_0^*(\bs{k}) & 0 & h_3^*(\bs{k}) & 0 & 0 & 0 & 0 & 0 & 0 & 0\\
0 & 0 & 0 & 0 & 0 & 0 & 0 & h_3(\bs{k}) & 0 & h_0(\bs{k}) & 0 & h_3(\bs{k}) & 0 & 0 & 0 & 0\\
0 & 0 & 0 & 0 & 0 & 0 & \gamma_1 & 0 & h_0^*(\bs{k}) & 0 & \gamma_1 & 0 & 0 & 0 & 0 & 0\\
0 & 0 & 0 & 0 & 0 & 0 & 0 & 0 & 0 & \gamma_1 & 0 & h_0(\bs{k}) & 0 & \gamma_1 & 0 & 0\\
0 & 0 & 0 & 0 & 0 & 0 & 0 & 0 & h_3^*(\bs{k}) & 0 & h_0^*(\bs{k}) & 0 & h_3^*(\bs{k}) & 0 & 0 & 0\\
0 & 0 & 0 & 0 & 0 & 0 & 0 & 0 & 0 & 0 & 0 & h_3(\bs{k}) & 0 & h_0(\bs{k}) & 0 & h_3(\bs{k})\\
0 & 0 & 0 & 0 & 0 & 0 & 0 & 0 & 0 & 0 & \gamma_1 & 0 & h_0^*(\bs{k}) & 0 & \gamma_1 & 0\\
0 & 0 & 0 & 0 & 0 & 0 & 0 & 0 & 0 & 0 & 0 & 0 & 0 & \gamma_1 & 0 & h_0(\bs{k})\\
0 & 0 & 0 & 0 & 0 & 0 & 0 & 0 & 0 & 0 & 0 & 0 & h_3^*(\bs{k}) & 0 & h_0^*(\bs{k}) & 0\\
\end{pmatrix},
\end{equation}
where $^*$ denotes the complex conjugate. The Hamiltonians for 3 and 4 layers are obtained by taking only the first 8 and 10 rows and columns respectively.

\section{Qualitative estimate of the broadening parameter $\delta$} 
\label{App:Broadening}

Assuming, for example, that the broadening originates from random disorder, we could use a simple approach to scattering and use the Born approximation:
$$
\delta \sim \mathrm{DOS} \times V,
$$
where $\mathrm{DOS}$ is the density of states calculated for the surface of graphite or for the bulk, and $V$ denotes the impurity potential amplitude. The bulk of graphite is three-dimensional and the electrons are approximately linearly dispersed, and hence the density of states is given by 
$$
\mathrm{DOS}_{bulk} = \frac{E^2}{2\pi^2\hbar^3 v^3},
$$
where $v$ is the velocity of the linearly dispersed electrons and $E$ is the energy at which we compute the DOS. At the surface, as pointed out in the main text, we have parabolically dispersed surface bands, and in two dimensions one has 
$$
\mathrm{DOS}_{surf} = \frac{m}{\pi \hbar^2}.
$$
Hence we have at $E = 0.01\,$eV:
$$
\frac{\delta_{bulk}}{\delta_{surf}} \sim \frac{1}{2\pi \hbar}\frac{E^2}{ m v^3} \frac{V_{bulk}}{V_{surf}} \sim \frac{1}{\pi}\frac{E^2 E_0}{ \hbar^3 v^3 k_0^2} \times 100 d_0 \approx 10,
$$
where we took $E_0 = 0.05\,$eV, $k_0 = 0.03\,$\AA$^{-1}$, $d_0 = 3.35\,$\AA$^{-1}$, and $v = 10^{15}\,$\AA/s. Above we assumed that we are dealing with a 100-layer graphite, and $k_0, E_0$ were taken from Fig.~\ref{fig:graphiteBernalKpoint} to make an estimate for the effective mass $m$ of the parabolic surface bands. 

\section{Band structure of bilayer graphene} 
\label{App:BLGbands}

\begin{figure}
	\centering
	\includegraphics[width=0.49\linewidth]{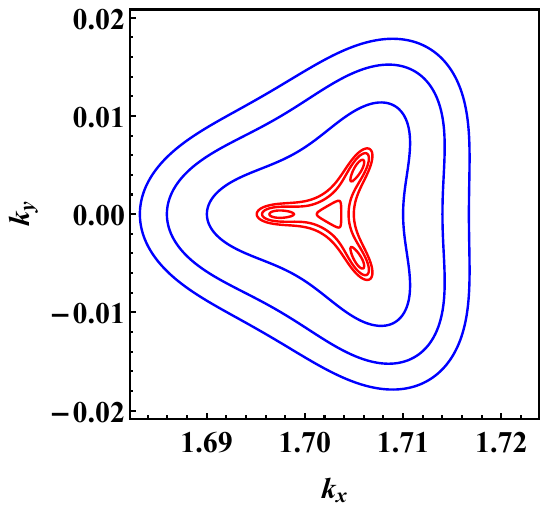}
		\includegraphics[width=0.49\linewidth]{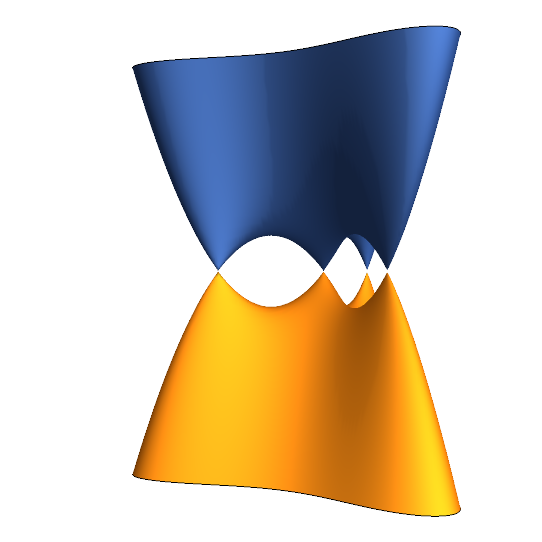}
	\caption{Left: Equal-energy contours of bilayer graphene electronic bands plotted for lower and higher energies (red and blue contours, respectively). Right: bands of bilayer graphene. We take $a_0 = 2.46\,$\AA, $\gamma_0 = 3.3\,$eV, $\gamma_1 = 0.42\,$eV, $\gamma_3 = -0.3\,$eV. The lower-energy contours are calculated at $E = 0.0007$, $0.0014$, and $0.002\,$eV, while the higher-energy contours are at $E = 0.01$, $0.02$, and $0.03\,$eV.}
	\label{fig:BLGbands}
\end{figure}

In Fig.~\ref{fig:BLGbands} we plot equal-energy contours and bands for bilayer graphene calculated from the Hamiltonian in Eq.~(\ref{hh1}). The red contours correspond to energies closer to the Dirac point split by the trigonal warping into four Dirac points. It is clear that the trigonal warping is better pronounced at lower energies. 

\end{document}